\documentclass[fleqn,usenatbib]{mnras}

\usepackage{newtxtext,newtxmath}

\usepackage[T1]{fontenc}

\usepackage{graphicx}	
\usepackage{amsmath}	
\usepackage{amssymb}	
\usepackage{comment}
\usepackage{booktabs}
\usepackage{xcolor}
\usepackage{soul}

\newcommand\crule[3][black]{\textcolor{#1}{\rule{#2}{#3}}}
\newcommand\Msun{\ensuremath{M_\odot}}

\title[Systematics in constraints from stellar streams]{Orbital phase-driven biases in Galactic mass constraints from stellar streams}

\author[Reino et al.]{
Stella Reino,$^{1}$
Robyn E. Sanderson,$^{2,3}$
Nondh Panithanpaisal,$^{2}$
Elena M. Rossi,$^{1}$
Konrad Kuijken$^{1}$
\\
$^{1}$Leiden Observatory, Leiden University, Niels Bohrweg 2, 2333 CA Leiden, The Netherlands\\
$^{2}$Department of Physics and Astronomy, University of Pennsylvania, 209 S 33rd St., Philadelphia, PA 19104, USA\\
$^{3}$Center for Computational Astrophysics, Flatiron Institute, 162 5th Ave., New York, NY 10010, USA\\
}

\date{Accepted XXX. Received YYY; in original form ZZZ}

\pubyear{2021}

\begin{document}
\label{firstpage}
\pagerange{\pageref{firstpage}--\pageref{lastpage}}
\maketitle

\begin{abstract}
One of the most promising tracers of the Galactic potential in the halo region are stellar streams. However, individual stream fits can be limited by systematic biases. To study these individual stream systematics, we fit streams in Milky Way-like galaxies from FIRE cosmological galaxy formation simulations with an analytic gravitational potential by maximizing the clustering of stream stars in action space. We show that for coherent streams the quality of the constraints depends on the orbital phase of the observed stream stars, despite the fact that the phase information is discarded in action-clustering methods. Streams on intermediate phases give the most accurate results, whereas pericentre streams can be highly biased. This behaviour is tied to the amount of correlation present between positions and momenta in each stream's data: weak correlation in pericentre streams prohibits efficient differentiation between potentials, while strong correlation in intermediate streams promotes it. 
Although simultaneous fitting of multiple streams is generally prescribed as the remedy to combat individual stream biases, we find that combining multiple pericentric streams is not enough to yield a bias-free result.
We finally show that adopting the two-component St\"ackel model does not fundamentally induce a biased mass estimate. With our full data set of two multi-wrap streams, we recovered the true rotation curve of the simulated galaxy within $12\%$ over the entire range of radii covered by our set of stars (10 - 176 kpc) and within $6.5\%$ between the 5 and 95-percentile distance range (23 - 109 kpc).
\end{abstract}



\section{Introduction}

Stellar streams, relics of tidally disrupted globular clusters and dwarf galaxies, are one of the most promising probes of the gravitational field of the Milky Way at large galactocentric distances. 
Since the first detections of streams in the Milky Way \citep{Ibata1994, Grillmair1995, Helmi1999}, many methods have been developed to constrain the Galactic mass profile using streams. Some of these make comparisons between predictions and data directly in position and velocity space such as the orbit-fitting technique \citep[e.g.][]{Koposov2010}, particle ejection methods (e.g. the "streakline" method of \citealp{Kupper2012}, the modified Lagrange Cloud Stripping method of \citealp{Gibbons2014}, the "particle-spray" method of \citealp{Fardal2015}) or full N-body simulations \citep[e.g.][]{LawMajewski2010}, while others utilize the action-angle coordinates, such as the angle-frequency slope method \citep{SandersBinney2013} and action-clustering method \citep{Sanderson2015}. 
These methods have thus far been applied to only a handful of streams, most commonly GD-1 \citep{Koposov2010, Malhan2019}, Pal 5 \citep{Kupper2015}, Sagittarius \citep{LawMajewski2010, Dierickx2017, Vasiliev2021} and Orphan \citep{Newberg2010, Erkal2019}.

To date, most of these studies have focused on measuring the potential with single streams (although see \citealp{Bovy2016}). However, using a sample of streams evolved in the Via Lactea II simulation \cite{Bonaca2014} showed  that constraints from individual streams can be highly biased. Only simultaneous fitting of a collection of streams would result in accurate potential recovery. They estimated that GD-1 and Pal 5-like streams could individually return up to $50\%$ biased mass estimates for the Milky Way. 
In \cite{Reino2021} we showed that this also holds true when using the action-clustering method. In particular, we saw a significant bias arise from the analysis of GD-1 compared to the constraints derived with a collection of streams. The strategy of simultaneous fitting of multiple streams to avoid the pitfalls of a single stream fit was also advocated by \cite{Lux2013}, \cite{SandersBinney2013} and \cite{Sanderson2017}.

Today, more than 60 streams have been discovered in the Milky Way \citep{NewbergCarlin2016, Mateu2018, Ibata2019, Myeong2019, Koppelman2019,  Naidu2020, Malhan2021} and ideally we should seek a consensus fit. However, the observational data required for Galactic potential inference is only available for a subset of these streams. Attaining the full 6D phase space information tends to be difficult and necessitates a cross-matching of information from different surveys, e.g. proper motions from Gaia \citep{Gaia2016}, distances of RR Lyrae from PanSTARRS1 survey \citep{Sesar2017} or Gaia's Specific Object Study catalogue \citep{Clementini2019} and radial velocities from RAVE \citep{RAVE2017}, WEAVE \citep{weave2012}, 4MOST \cite{4most} or DESI \citep{DESI2019}. Since steams are typically distant and faint, targeted follow-up surveys are often needed, such as the H3 survey \citep{H32019} targeting the stellar halo and the S5 survey \citep{S52019} targeting the stellar streams in the Southern Hemisphere. With this in mind, knowledge of which streams are the most useful for putting accurate constraints on the Galactic potential would be valuable for selecting which streams to focus both our modelling and observing efforts on.

To gain this insight, \cite{Bonaca2018} explored the intrinsic information content in the tracks of 11 mock globular cluster stellar streams as a function of their properties using the Fisher-matrix approach. They found that angular length of the stream was the best predictor of the tightness of their parameter constraints and that streams on more eccentric orbits were the most sensitive to the halo shape. However, while the Fisher-matrix approach allowed them to investigate the precision of the constraints the different streams were capable of reaching, they did not address the question of accuracy. 

In this paper, we aim instead to gauge the accuracy of the constraints that arise from different streams using the action-clustering method. In particular, we explore the systematics of stream-driven constraints as a function of their orbital phase. To this end, we select two long spatially coherent streams from FIRE cosmological-hydrodynamical simulations \citep{Hopkins2018} for our analysis. First, we set the expectation for the accuracy obtainable with two full-length streams and then divide the streams into smaller sections based on their current orbital phase and analyse these segments independently.
This approach is motivated, first, by the fact that in reality it is more likely to detect (or have the full 6D phase space information for) only a small nearby segment of the whole longer stream, or that associations between stream sections observed in different parts of the sky are uncertain. Second, the streams most commonly used for characterising the potential of the Milky Way are from globular cluster origin, and therefore much shorter than the dwarf galaxy streams found in FIRE simulations. Since globular cluster streams do not intrinsically form in the FIRE simulations, we can instead approximate their length by treating each section as an independent stream. Third, this approach allows us to keep constant some properties of the streams we are working with (for instance eccentricity, inclination, apocentre distance), and to direct our focus on phase differences. Finally, with this framework we can easily apply our method to the full stream data to verify that our potential model does not induce severe systematic biases on these sectional results.
As a by-product, this setup will give us an indication of whether, and how much, the constraints vary over the length of the full long stream.
Another goal of our study is to investigate how accurately we can recover the true potential of the simulated galaxy when modelling the streams with a St\"ackel potential. Although St\"ackel potentials are generally considered inadequate for describing realistic galaxies, they have the great benefit of exact actions. This property was our incentive for adopting the St\"ackel potential when analysing real Milky Way stellar streams in \cite{Reino2021}. In this work, we will test whether this assumption can introduce any significant additional bias into our results.

This paper is organised as follows. We discuss the elements of our method in Section~\ref{sec:method}, introducing the St\"ackel potential (\ref{sec:St\"ackel}), the action coordinates (\ref{sec:actions}) and our action-space clustering measure (\ref{sec:kld}). We give an overview of the FIRE suite of simulations and our stream sample in Section~\ref{sec:data}. In Section~\ref{sec:fullstreams} we present our results for the full streams, while the results for stream sections are shown in Section~\ref{sec:sections}. Next, in Section~\ref{sec:phase} we explore the stream section results as a function of orbital phase. In Section~\ref{sec:residuals} we explore the dependence of our result on other stream section properties and provide a reasoning for the orbital-phase effects. Finally, we discuss our results and make conclusions in Section~\ref{sec:discussion}.

\section{Method}
\label{sec:method}

We follow the action-clustering method outlined in \cite{Sanderson2015} and \cite{Reino2021} which aims to constrain the galactic potential by maximizing the clustering of stream stars in action space. We vary the potential used for converting the stars’ positions and velocities into action coordinates and adopt as the best-fit potential the one that gives rise to the most clustered distribution of actions.

\subsection{St\"ackel potential}
\label{sec:St\"ackel}

Analytical transformation of phase space coordinates $(\boldsymbol{x}, \boldsymbol{v})$ to action coordinates $\boldsymbol{J}$ is possible only for a small set of potentials for which Hamilton-Jacobi equations can be solved by separation of variables. Action estimation for general potentials requires the use of numerical approximation algorithms \citep[see][]{Sanders2016}, although repeated calculation of actions can be computationally costly in these cases.
Analytical calculation, in contrast, is less expensive and as such allows both for a larger number of stream stars to be included in the fit and more potentials to be considered.

The St\"ackel potential, which separates in ellipsoidal coordinates, is the most general of the small group of separable potentials to describe a real galaxy, as it allows for arbitrarily flattened density profiles and construction of flat rotation curves with the two-component St\"ackel model \citep{Batsleer1994}. 

However, the applicability of St\"ackel potentials is limited by the fact that all orbits are defined by the same foci. This restriction has been shown to be incorrect for real galaxies \citep{Binney2012, KuijkenGilmore1989} and therefore a perfect global fit is not possible. Many action-calculation methods still utilize its advantages and, for example, fit each orbit with a local St\"ackel potential instead \citep{Sanders2012} or apply the St\"ackel formulae to another more complex potential of interest \citep{Binney2012}.

One of the aims of the current work is to test the practicality of adopting a St\"ackel potential to describe a realistic galaxy and to see whether, despite its known limits, it could still be utilized as a valuable tool. Discussion on the expected size of errors originating from our choice of the St\"ackel potential model is included in Section~\ref{sec:fullstreams}.

In this work, we consider oblate axisymmetric two-component St\"ackel potentials described by spheroidal coordinates, a limiting case of ellipsoidal coordinates.
Spheroidal coordinates $(\lambda, \nu, \phi)$ are related to cylindrical coordinates $(R, z,\phi)$ by the following quadratic equation
\begin{equation}
\label{eq:transformation}
\frac{R^2}{\tau -a^2} + \frac{z^2}{\tau -c^2} = 1 \ ,
\end{equation}
where $\tau = \lambda, \nu$ are the roots.

Parameters $a$ and $c$ are constants that determine the location of the foci $\Delta = \sqrt{a^2 - c^2}$ and define the axis ratio of the coordinate surfaces, $e \equiv \frac{a}{c}$ and, therefore, establish the shape of the coordinate system. For an oblate density distribution we have $a > c$. 
Further details about this coordinate system can be found in \cite{deZeeuw1985} and \cite{DZ1988}.

A St\"ackel potential, $\Phi$, in spheroidal coordinates can be written as
\begin{equation}
\begin{aligned}
\label{eq:St\"ackel}
&\Phi (\lambda, \nu) = - \frac{f(\lambda) - f(\nu)}{\lambda -\nu} \ ,\\
&f(\tau) = (\tau - c^2) \mathcal{G} (\tau) \ ,
\end{aligned}
\end{equation}
where we set $\mathcal{G}(\tau)$ to be the Kuzmin-Kutuzov potential
\begin{equation}
\mathcal{G} (\tau) = \frac{GM_{\mathrm{tot}}}{\sqrt{\tau} + c} \ ,
\end{equation}
with $M_{\mathrm{tot}}$ the total mass and G the gravitational constant.

To construct a two-component St\"ackel model we combine two individual St\"ackel potentials, $\Phi_{\mathrm{outer}}$ and $\Phi_{\mathrm{inner}}$. The motivation for this is to add more flexibility to our potential model thereby allowing for a more realistic model of the galaxy \citep{Batsleer1994, Reino2021}.
The two components are defined by different parameters $a_{\mathrm{outer}}$, $c_{\mathrm{outer}}$ and $a_{\mathrm{inner}}$, $c_{\mathrm{inner}}$, and therefore each has a different scale and axis ratio but, crucially, they must have the same foci for the total potential to retain the St\"ackel form (as defined by Equation~\ref{eq:St\"ackel}). It then follows that
\begin{equation}
\begin{aligned}
\label{eq:coordlinks}
&a_{\mathrm{outer}}^2 - a_{\mathrm{inner}}^2 = c_{\mathrm{outer}}^2 - c_{\mathrm{inner}}^2 = q \ ,\\
&\lambda_{\mathrm{outer}} - \lambda_{\mathrm{inner}} = \nu_{\mathrm{outer}} - \nu_{\mathrm{inner}} = q \ , 
\end{aligned}
\end{equation}
where $q>0$ is a constant. The total $\mathcal{G} (\tau)$ is now a combination of two parts, $\mathcal{G}_{\mathrm{outer}} (\tau)$ and  $\mathcal{G}_{\mathrm{inner}} (\tau)$
\begin{equation}
\mathcal{G} (\tau) = \frac{GM_{\mathrm{tot}}(1-k)}{\sqrt{\tau_{\mathrm{outer}}} + c_\mathrm{outer}} +  \frac{GM_{\mathrm{tot}}k}{\sqrt{\tau_\mathrm{outer}- q} + c_{\mathrm{inner}}}\ ,
\end{equation}
and the total two-component potential is
\begin{equation}
\begin{aligned}
&\Phi (\lambda_{\mathrm{outer}}, \nu_{\mathrm{outer}}, q) =  \\
&-GM_{\mathrm{tot}}\Bigg[ \frac{1-k}{\sqrt{\lambda_{\mathrm{outer}}} + \sqrt{\nu_{\mathrm{outer}}}} + \frac{k}{\sqrt{\lambda_{\mathrm{outer}}-q} + \sqrt{\nu_{\mathrm{outer}}}}\Bigg]
\end{aligned}
\end{equation}
where $k$ is the ratio between the inner component and the outer component masses and $M_{\rm tot}$ is the sum of the two component masses.

We define our two-component St\"ackel potentials on a grid of five parameters $\boldsymbol{\zeta} = (M_{\mathrm{tot}}, a_{\mathrm{outer}}, e_{\mathrm{outer}}, a_{\mathrm{inner}}, k)$.
We select the trial potentials by drawing $50$ points for each of the shape parameters, from uniform distributions in log space, over the ranges: [0.7, 1.8] in $ \log_{10}(a_{\mathrm{outer}}/ \rm kpc)$, [$\log_{10}(1.0), \log_{10}(2.0)$] in $\log_{10}(e_{\mathrm{outer}})$ and [$\log_{10}(0.), \log_{10}(0.7)$] in $\log_{10}(a_\mathrm{inner}/ \rm kpc)$. However, we only use a subset of these parameter combinations ($\sim 8000$) that constructs a mathematically valid potential according to the equation~\ref{eq:coordlinks}; i.e., the parameter combinations which adhere to $c_{\mathrm{outer}}^2 = \frac{e_{\mathrm{outer}}^2}{a_{\mathrm{outer}}^2} > c_{\mathrm{inner}}^2$.
We also draw $20$ points for each mass parameter over the ranges: [11.5, 12.5] in $\log_{10}(M/M_{\odot})$ and [$\log_{10}(0.01), \log_{10}(0.3)$] in $\log_{10}(k)$. In total, our grid contains $3\, 253\, 600$ trial potentials.

We further discard the potentials that cause any of the star particles in our sample to be unbound from the host galaxy. In \cite{Reino2021}, we showed that our results did not change appreciably if, instead of this strict condition, we allowed a small percentage of the stars to become unbound. Furthermore, in that work we were dealing with real stream data with measurement errors while here we know the true position and velocity of all our particles and do not need to worry about measurement errors causing unbound stars.

\subsection{Actions}
\label{sec:actions}

The action-angle coordinates are a set of canonical coordinates
which considerably simplify the equations of motion of a bound star in a static or adiabatically time-evolving potential: the actions, $J_i$, are integrals of motion that uniquely define the stellar orbit and the angles, $\theta_i$, are periodic coordinates that express the phase of the orbit.
For the St\"ackel potential we define the actions $J_{\lambda}$ and $J_{\nu}$ as
\begin{equation}
\label{eq:st_actions}
J_{\tau} = \frac{1}{2 \pi} \oint p_{\tau} d\tau \ ,
\end{equation}
where $p_{\tau}$ is the conjugate momentum to the coordinate $\tau$ and the integral is over the full oscillation in $\tau$. The third action $J_{\phi}$ is equal to the z-component of the angular
momentum, $L_z$, and hence is constant in our axisymmetric potentials.

The conjugate momenta, $p_{\tau}$, can be found by solving the Hamilton-Jacobi equation by separation of variables. In addition to the momenta, the separation of variables introduces three isolating integrals: $I_2$, $I_3$ and the total energy $E$.
The integrals $I_2$ and $I_3$ are defined as \citep{DZ1988}
\begin{equation}
\begin{aligned}
&I_2 = \frac{L_z^2}{2} ,\\
&I_3 = \frac{1}{2} (L_x^2 + L_y^2) + (a^2 - c^2) \Big[ \frac{1}{2} v_z^2 - z^2 \frac{\mathcal{G} (\lambda) - \mathcal{G} (\nu)}{\lambda - \nu} \Big] \ .
\end{aligned}
\end{equation}
The solution to the Hamilton-Jacobi equation then allows the momenta, $p_{\tau}$, to be expressed as a function of the $\tau$ coordinate and the three isolating integrals:
\begin{equation}
p_{\tau}^2 = \frac{1}{2 (\tau - a^2)} \Big[ \mathcal{G} (\tau)  - \frac{I_2}{\tau - a^2} - \frac{I_3}{\tau - c^2} + E \Big] \ ,
\end{equation}
which can then be used in Equation~\ref{eq:st_actions} and integrated numerically.

\subsection{Clustering measurement}
\label{sec:kld}

An intrinsic element of the action-clustering method is the procedure of quantifying and comparing the degree of clustering present in the action space of different potentials. Following \cite{Sanderson2015} and \cite{Reino2021}, we measure this degree of clustering with the
Kullback-Leibler divergence (KLD). The KLD is a measure of the divergence between two probability distributions $p(\boldsymbol{x})$ and $q(\boldsymbol{x})$.
For a discrete sample $[\boldsymbol{x}_i]$ with $\ i = 1,...\, ,N$ drawn from a distribution $p(\boldsymbol{x})$, the KLD can be calculated as
\begin{equation}
\mathrm{KLD} (p \ ||\ q) \approx \frac{1}{N} \sum_i^N \log \frac{p(\boldsymbol{x}_i)}{q(\boldsymbol{x}_i)}, \quad \mathrm{if} \ q(\boldsymbol{x}_i) \neq 0 \ \forall i.
\end{equation}
The value of KLD increases with increasing difference between $p(\boldsymbol{x})$ and $q(\boldsymbol{x})$ and is equal to 0 when $p(\boldsymbol{x}) = q(\boldsymbol{x})$.

Since we are looking to measure the amount of clustering in action space, irrespective of the cluster locations, we proceed with the idea of comparing the distribution of actions in a particular trial potential to a completely unclustered, featureless distribution. 
In other words, we set $q(\boldsymbol{x})$ to a uniform distribution in the actions, $u(\boldsymbol{J})$, and $p(\boldsymbol{x})$ to a probability distribution of actions, $p(\boldsymbol{J} \mid \boldsymbol{\zeta}, \boldsymbol{\omega})$, found by mapping the phase space coordinates $\boldsymbol{\omega}$ to action space $\boldsymbol{J}$ with a trial potential parameterized by $\boldsymbol{\zeta}$.
The difference between the two distributions is greater the more clustered the action distribution is.

Our goal is to maximize the difference, and therefore the KLD value, between these probability distributions as we vary the trial potential and explore the parameter space. The trial potential with the highest KLD value is adopted as the best-fit potential, with parameters $\boldsymbol{\zeta}_0$, for that particular data set, $\boldsymbol{\omega}$.

The standard KLD gives equal weight to each star particle in the sample and as a consequence, when multiple streams (or stream sections) are analysed simultaneously, streams (or stream sections) containing more star particles have stronger influence over the results. Since stream membership in this case is known we can make the best use of the data by giving equal weight to all \emph{streams} instead, by weighting the contribution of each star particle with
\begin{equation}
\label{eq:w}
w_{j} = \frac{1}{N_{\rm s}} \times \frac{1}{N_{j}} \ ,
\end{equation}
where $N_{\rm s}$ is the number of streams and $N_{j}$ is the number of star particles in stream $j$.

This {\em weighted} KLD is thus calculated as follows:
\begin{equation}
\label{eq:kld1_w}
\mathrm{KLD1}(\boldsymbol{\zeta}) = \sum_j^{N_{\rm s}} \sum_i^{N_{j}} \ w_j \log \frac{p(\boldsymbol{J} \mid \boldsymbol{\zeta}, \boldsymbol{\omega})}{u(\boldsymbol{J})} \biggr\rvert_{\boldsymbol{J} = \boldsymbol{J}_{\zeta}^{ij}}\ ,
\end{equation}
where $\boldsymbol{J}_{\zeta}^{ij} = \boldsymbol{J} (\boldsymbol{\zeta}, \boldsymbol{\omega}_{ij})$ and $\boldsymbol{\omega}_{ij}$ are the phase space coordinates for star i in stream j. 

Although the function $p(\boldsymbol{J} \mid \boldsymbol{\zeta}, \boldsymbol{\omega})$ is not known a priori, it can be constructed using the observed points $\boldsymbol{J}$ via a density estimator. Here, we obtain $p(\boldsymbol{J} \mid \boldsymbol{\zeta}, \boldsymbol{\omega})$ using the Enlink algorithm developed by \cite{Sharma2009}. $u(\boldsymbol{J})$, on the other hand, is constant across all trial potentials and doesn't have an impact on the results. It can, therefore, be set to any preferred value.

Finally, the method will work best if the different streams do not overlap with each other in action-space. Since in this case, as in \cite{Reino2021}, we know which stars belong to which stream, we can ensure this absence of overlap by calculating the probability distributions for each stream independently. So, instead of estimating $p(\boldsymbol{J} \mid \boldsymbol{\zeta}, \boldsymbol{\omega})$ with the full set of sample points $\boldsymbol{J}$, we construct a probability density function $p_j(\boldsymbol{J}_j \mid \boldsymbol{\zeta}, \boldsymbol{\omega}_j)$ for each stream j individually from points $\boldsymbol{J}_j$. To keep the $p_j$ at the correct relative size between the different streams, we normalize each $p_j$ with $\frac{N_j}{N}$. In practice, the KLD equation that we use is therefore
\begin{equation}
\label{eq:kld1}
\mathrm{KLD1}(\boldsymbol{\zeta}) = \sum_j^{N_{\rm s}} \sum_i^{N_{j}} \ w_j \log \frac{N_j}{N} \frac{p_j(\boldsymbol{J}_j \mid \boldsymbol{\zeta}, \boldsymbol{\omega}_j)}{u(\boldsymbol{J})} \biggr\rvert_{\boldsymbol{J} = \boldsymbol{J}_{\zeta}^{ij}}\ .
\end{equation}

It is important to note that neither the weighing nor the separation of streams in action-space is critical for the action-clustering method to work, as we have shown in previous works \citep{Sanderson2015, Reino2021}.

The goal of this procedure, as already stated above, is to determine the set of St\"ackel potential parameters that maximize the KLD1 value.
As a second step in our procedure, we calculate the confidence intervals on these best-fit parameters, by comparing the action distribution of the best-fit potential, $p(\boldsymbol{J} \mid \boldsymbol{\zeta}_0 \,, \boldsymbol{\omega})$, to the action distributions of the other trial potentials, $p(\boldsymbol{J} \mid \boldsymbol{\zeta}_{\mathrm{trial}} \,, \boldsymbol{\omega})$ using again the Kullback-Leibler divergence. 

The KLD defined in such a way can be interpreted as the relative probability of the potential parameters $\boldsymbol{\zeta}_{\mathrm{trial}}$ to the best-fit potential parameters $\boldsymbol{\zeta}_0$ \citep{Kullback1951, Kullback1959}. As we move further from the best-fit parameter values, the difference between the two action distributions grows and so does the KLD value. In other words, we can measure how far from the best-fit parameters we can move before the action distribution starts to significantly differ from that of the best-fit action-distribution. The confidence intervals can then be drawn based on the value of KLD that we deem to correspond to significant difference. The interpretation of KLD as the expectation value of the difference in the log of two posterior probabilities allows us to determine the value of KLD that corresponds to any preferred level of significance. For example, the significance of $1\sigma$ corresponds to KLD = 0.5. An in-depth derivation of the relation between the KLD values and the confidence levels can be found in \cite{Reino2021} and a full discussion of this interpretation of KLD in \cite{Sanderson2015}.

As with $\mathrm{KLD1}(\boldsymbol{\zeta})$ in equation~\ref{eq:kld1}, this version of KLD will incorporate weights and a separate density estimation for different streams. It is defined as follows
\begin{equation}
\label{eq:kld2}
\mathrm{KLD2}(\boldsymbol{\zeta}) = \sum_j^{N_s} \sum_i^{N_j} \ w_j \log \frac{N_j}{N} \frac{p_i(\boldsymbol{J}_i \mid \boldsymbol{\zeta}_0, \boldsymbol{\omega}_i)}{p_i(\boldsymbol{J}_i \mid \boldsymbol{\zeta}_{\mathrm{trial}}, \boldsymbol{\omega}_i)} \biggr\rvert_{\boldsymbol{J}_i = \boldsymbol{J}_0^{ij}}\ .
\end{equation}
Once again, we use Enlink to estimate the probability density functions $p_i(\boldsymbol{J}_i \mid \boldsymbol{\zeta}_0, \boldsymbol{\omega}_i)$ and $p_i(\boldsymbol{J}_i \mid \boldsymbol{\zeta}_{trial}, \boldsymbol{\omega}_i)$ using the two sets of actions. Both functions are then evaluated at $\boldsymbol{J}_0 = \boldsymbol{J}(\boldsymbol{\zeta}_0 \,, \boldsymbol{\omega})$, the actions computed with the best-fit potential parameters $\boldsymbol{\zeta_0}$. 

Finally, throughout the rest of the paper we discuss the uncertainties of our measurements as $1\sigma$ confidence intervals. This corresponds to the subset of trial potentials with $\mathrm{KLD2}(\boldsymbol{\zeta}) \leq 0.5$. The individual parameter confidence intervals are determined as the full range of parameter values in this subset of potentials.

\begin{figure*}
\centering
\includegraphics[width=0.97\linewidth, trim={0 10 0 10}, clip]{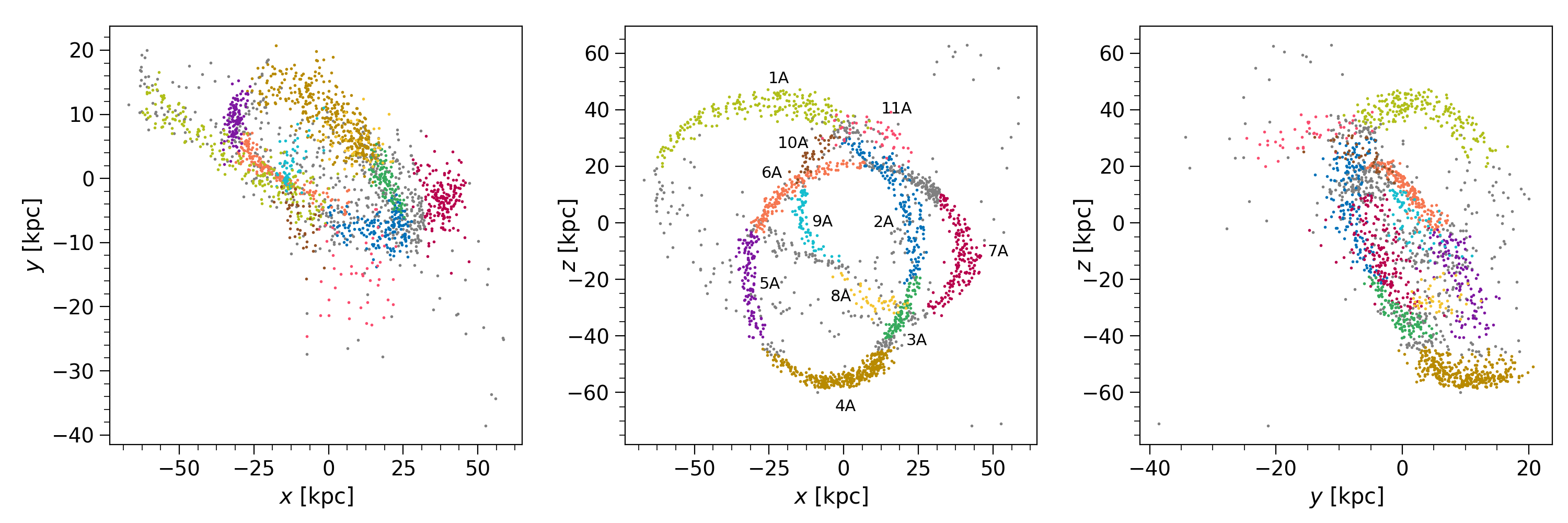}\
\includegraphics[width=0.97\linewidth, trim={0 10 0 10}, clip]{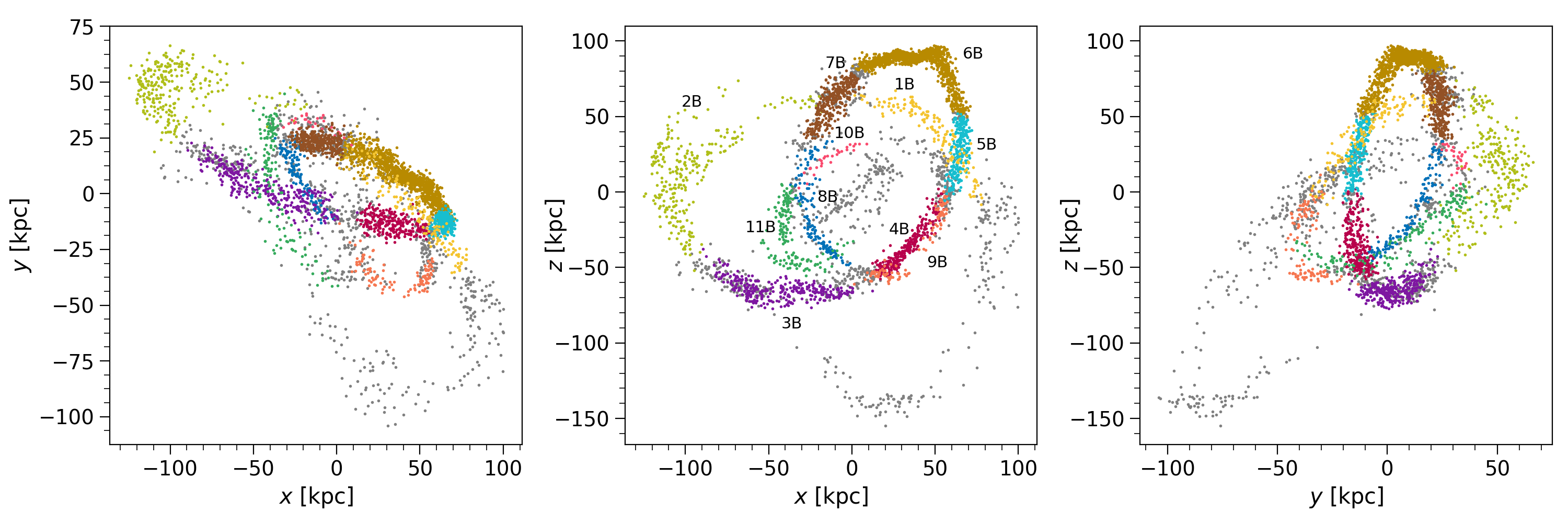}
\caption{Streams A (top) and B (bottom) in galactocentric coordinates. Each stream section is identified by a given colour throughout the whole paper (see Table \ref{tab:sections} for a summary). Star particles not belonging to any section are shown as grey.}
\label{fig:streams}
\end{figure*}

\begin{figure*}
\centering
\includegraphics[width=0.49\linewidth, trim={0 10 0 10}, clip]{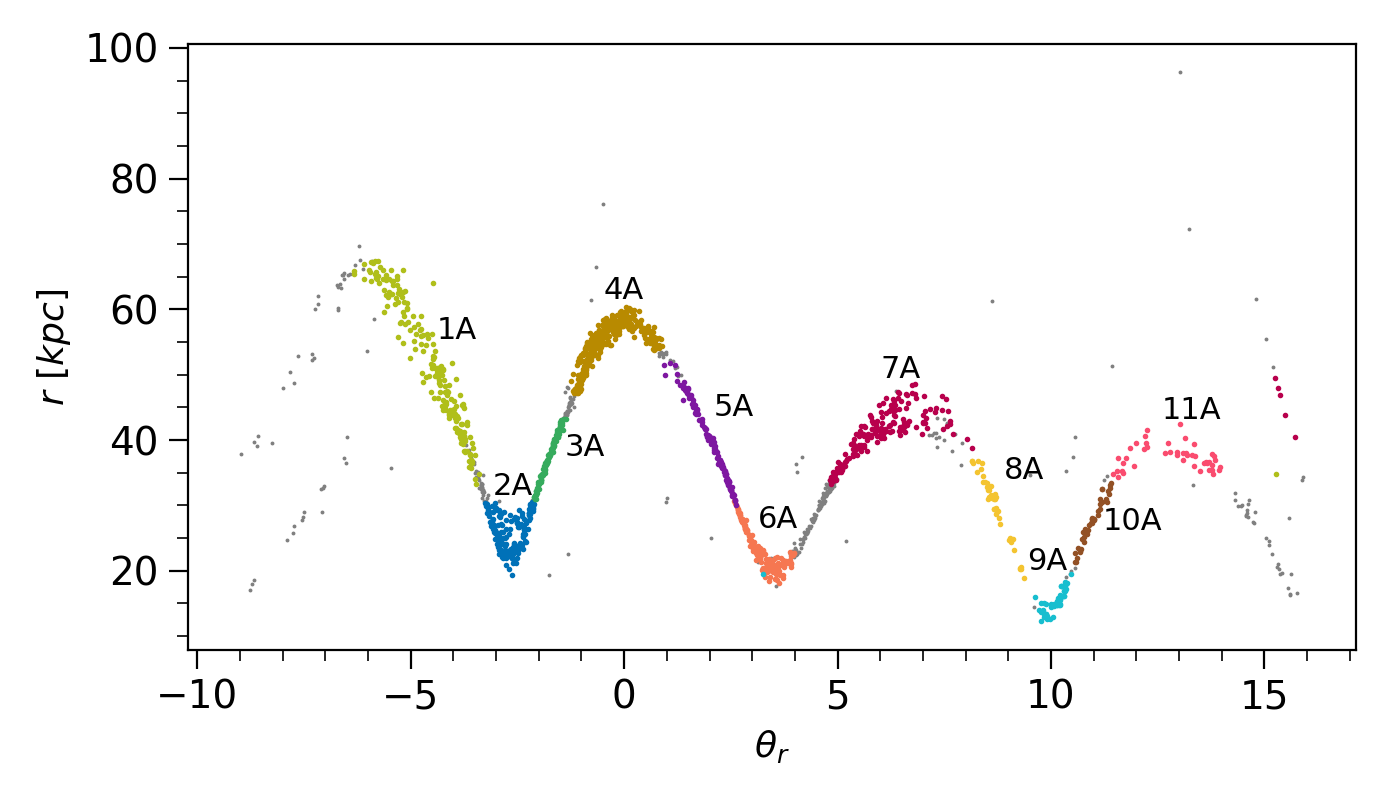}
\includegraphics[width=0.49\linewidth, trim={0 10 0 10}, clip]{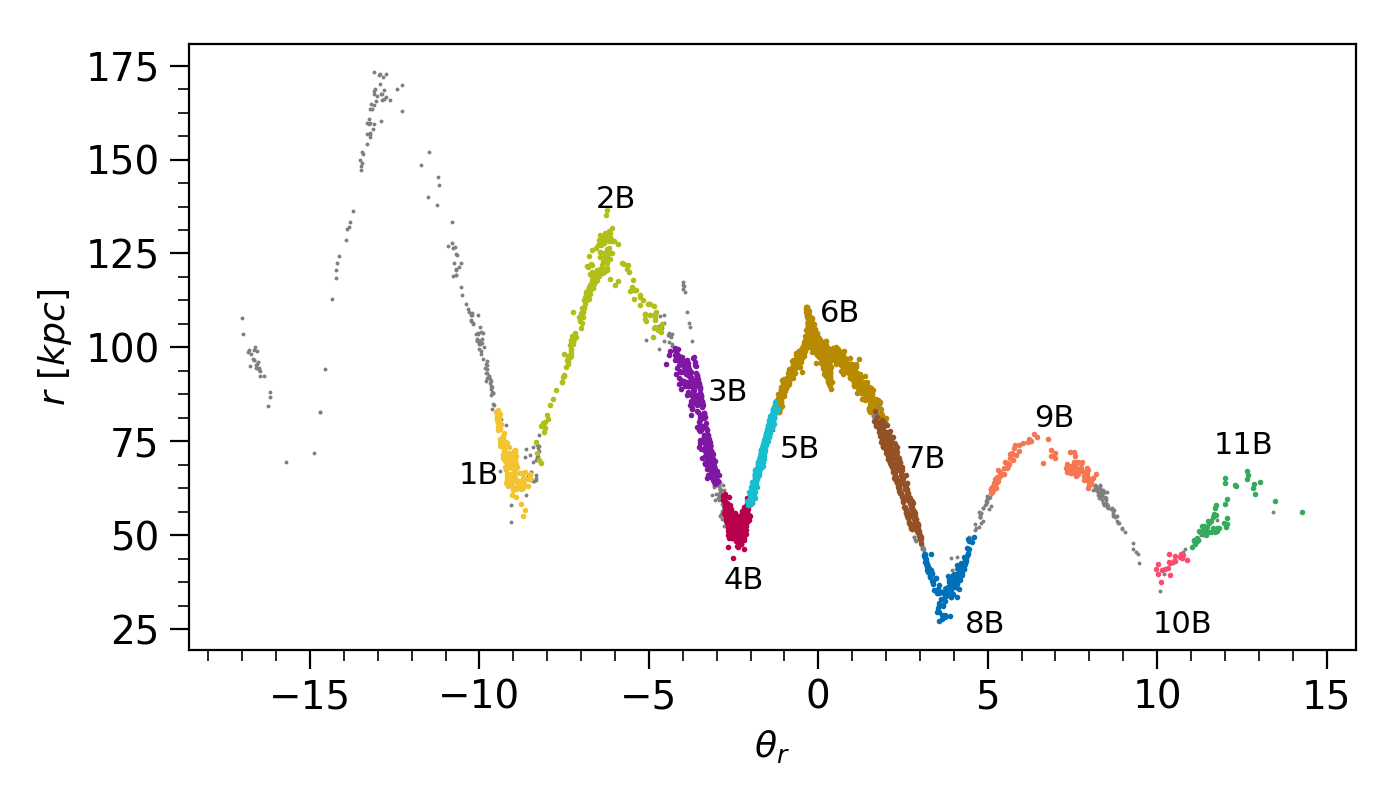}
\caption{Phase information of streams A (left) and B (right). The progenitor-containing section (in gold) is at $\theta_r = 0$. Colour scheme as in Figure~\ref{fig:streams} (see also Table~\ref{tab:sections}).}
\label{fig:phases}
\end{figure*}

\definecolor{french blue}{HTML}{0071B8}
\definecolor{amaranth purple}{HTML}{B8004C}
\definecolor{burnt sienna}{HTML}{F67751}
\definecolor{dark goldenrod}{HTML}{B88A00}
\definecolor{violet}{HTML}{7e17a1}
\definecolor{blue green}{HTML}{17BECF}
\definecolor{acid green}{HTML}{B0BF1A}
\definecolor{fiery rose}{HTML}{FA4C6F}
\definecolor{green pigment}{HTML}{36AB5D}
\definecolor{saddle brown}{HTML}{935125}
\definecolor{saffron}{HTML}{F4C430}

\begin{table*}
\begin{tabular}{ c c c c c c c c}
Stream & $\mathrm{N}_*$ & $r$ [kpc] & Phase & Colour & Length [kpc] & Width [kpc] & $\sigma_v$ [$\mathrm{km \ s^{-1}}$] \\
\hline
A & 2063 & 37.7 & - & - & - & - & - \\
\hline
1A & 215 & 50.8 & apocentre & \crule[acid green]{0.75cm}{0.1cm} & 74.7	& 3.63 &	31.9\\
2A & 165 & 26.4 & pericentre & \crule[french blue]{0.75cm}{0.1cm} & 61.6	& 2.60 & 20.8\\
3A & 109 & 38.6 & intermediate & \crule[green pigment]{0.75cm}{0.1cm} & 38.6 & 1.84 & 18.6\\
4A & 362 & 55.6 & apocentre (progenitor) & \crule[dark goldenrod]{0.75cm}{0.1cm} & 50.8 & 3.12 & 12.1\\
5A & 117 & 38.4 & intermediate & \crule[violet]{0.75cm}{0.1cm} & 33.5 & 2.69 & 14.5\\
6A & 167 & 23.3 & pericentre & \crule[burnt sienna]{0.75cm}{0.1cm} & 47.0 & 1.67 & 23.2 \\
7A & 183 & 41.5 & apocentre & \crule[amaranth purple]{0.75cm}{0.1cm} & 45.7 & 3.49 & 36.8\\
8A & 43 & 31.1 & intermediate & \crule[saffron]{0.75cm}{0.1cm} & 25.8 &	3.77 & 25.1\\
9A & 70 & 15.0 & pericentre & \crule[blue green]{0.75cm}{0.1cm} & 32.1 &	2.37 &	61.3\\
10A & 41 & 27.8 & intermediate & \crule[saddle brown]{0.75cm}{0.1cm} & 23.3 & 2.92 & 34.6\\
11A & 45 & 37.2 & apocentre & \crule[fiery rose]{0.75cm}{0.1cm} & 34.1 &	6.28 & 40.6 \\
\hline
B & 4038 & 78.8 & - & - & - & - & -\\
\hline
1B & 149 & 68.2 & pericentre & \crule[saffron]{0.75cm}{0.1cm} & 115.5 &	5.46 &	20.6\\
2B & 263 &  115.2 & apocentre & \crule[acid green]{0.75cm}{0.1cm} & 179.7 & 13.0 & 21.7\\
3B &  305 & 80.8 & intermediate & \crule[violet]{0.75cm}{0.1cm} & 110.9 &	6.48 &	16.4\\
4B & 259 & 53.5 & pericentre  & \crule[amaranth purple]{0.75cm}{0.1cm} & 69.0 &	4.39 &	16.1\\
5B & 243 & 72.4 & intermediate & \crule[blue green]{0.75cm}{0.1cm} & 58.4 & 3.66 & 12.0\\
6B & 1152 & 96.8 & apocentre (progenitor) & \crule[dark goldenrod]{0.75cm}{0.1cm} & 93.8 &	4.75 & 11.1\\
7B & 318 & 67.9 & intermediate & \crule[saddle brown]{0.75cm}{0.1cm} & 47.3 & 5.86 & 19.0\\
8B & 144 & 38.8 & pericentre & \crule[french blue]{0.75cm}{0.1cm} & 103.1 & 3.45 & 22.6\\
9B & 109 & 66.9 & apocentre  & \crule[burnt sienna]{0.75cm}{0.1cm} & 108.2 & 4.71 & 20.3\\
10B & 30 & 42.9 & pericentre & \crule[fiery rose]{0.75cm}{0.1cm} & 55.8 &	2.42 & 19.0\\
11B & 159 & 53.3 & apocentre & \crule[green pigment]{0.75cm}{0.1cm} & 118.1 &	6.44 & 24.7\\
\hline 
\end{tabular}
\caption{Sections defined in our streams. The upper portion lists the sections from stream A and the lower portion contains those of stream B, each with their respective full stream details on top. The columns give for each stream their signifier, number of stars $N_*$, median galactocentric distance $r$, approximate orbital phase, the colour scheme (which is used to mark the corresponding sections throughout the paper), length, width and velocity dispersion. Discussion on how these properties were computed is included in Appendix~\ref{sec:properties}.}
\label{tab:sections}
\end{table*}

\section{Simulation data}
\label{sec:data}

In this work, we make use of streams formed from the tidal disruption of dwarf galaxies in cosmological-baryonic simulations of Milky Way-like galaxies from the Latte suite \citep[][]{Wetzel2016} and ELVIS on FIRE suite \citep{GarrisonKimmel2019} of the Feedback In Realistic Environments (FIRE) project \citep[][]{Hopkins2018}. All halos were simulated in $\Lambda$CDM cosmology at particle mass resolution of 3500--7100 $\Msun$ and spatial resolution of 1--4 pc for star/gas particles; 18,000--35,000 $\Msun$ and 40 pc for DM particles. The resolution of this suite of simulations allows both luminous and dark subhalos to be resolved well even near each MW-like galaxy, and follows the formation of tidal streams from dwarf galaxies down to slightly below the mass of the MW's ``classical`` dSphs: around $10^8\ \Msun$ in total mass or $10^6\ \Msun$ in stellar mass (at $z=0$). 

\cite{Panithanpaisal2021} search these simulated galaxies for accreted structures that are spatially coherent and stream-like at present day. They identify 100 such streams across 13 simulations (see their Table 1) and confirm that the progenitor galaxies of these coherent streams are consistent with the mass-size-velocity dispersion relationship of observed present-day Milky Way satellites. This implies that the streams' phase-space volumes, and therefore their sizes and densities in action space, are representative of real streams from satellite galaxies. 

In the current work we focus on two of the nine coherent tidal streams found in the halo of the isolated galaxy simulation \textsf{m12i}. This simulated galaxy has had a quiet recent accretion history involving mostly quite low-mass galaxies, as the MW's is expected to have been since the Gaia-Enceladus merger (e.g. \citealt{Bonaca2017}, \citeyear{Bonaca2020}; \citealt{Belokurov2018}, \citeyear{Belokurov2020}; \citealt{Haywood2018}, \citealt{DiMatteo2019}, \citealt{Naidu2020}). Specifically, \textsf{m12i} experiences no mergers with mass ratios more similar than 1:3 after $z=1.7$ (about 9.5 Gyr ago; \citealt{2020MNRAS.497..747S}). Its thin disk stabilizes in its current configuration more than 5 Gyr before present day \citep{2018MNRAS.481.4133G}, and has a stellar mass and surface density at the Solar circle comparable to the MW \citep{2020ApJS..246....6S}. 

We select two long streams identified in \textsf{m12i}, which each have multiple wraps around their host galaxy and contain between 2000-4000 star particles. Although they are far longer than nearly every known MW stream, this length is ideal for our purposes as it allows us to divide each stream into many sections that are each comparable to most known stream lengths, and to select several instances of the same orbital phase from each stream. Throughout the paper, we use the error-free present day positions and velocities of the star particles that belong to these streams and assume complete and contamination-free knowledge of stream membership for each star particle. 

We define 11 sections in each stream by eye using a combination of the position, velocity and orbital phase information. Figure~\ref{fig:streams} shows the two streams in the galactocentric reference frame with each section highlighted using a different colour (the velocities of the streams are shown in Appendix~\ref{sec:stream_velocities} in Figure~\ref{fig:streams_v}). Throughout this paper, we consistently use the same colour to represent a particular section of each stream, the legend for this section-specific colour scheme is given in Table~\ref{tab:sections}. Gold-coloured points represent the progenitor-containing section in both streams. The star particles that don't belong to any section are shown in light grey. In Figure~\ref{fig:phases} we show the streams after we have unwound them using angle coordinates. The angle coordinates were computed using the AGAMA library \citep{Vasiliev2019} in a low-order multipole (dark matter and hot gas distribution) and cylindrical spline (stellar and cold gas distribution) model fit to the potential of \textsf{m12i} (Arora et al in prep). To unwind each stream, we utilized the Hough transform \citep{hough}, a line detection algorithm, to identify line overdensities in the $\theta_r$ vs. $\theta_\phi$ projection \citep{Shih2021, Pearson2021}. All the lines identified were then connected by exploiting the periodic boundary condition in the angle projection (i.e. if a line terminates at $(\theta_r, \theta_\phi) = (2\pi, \Tilde{\theta}_\phi)$, it will reappear at $(\theta_r, \theta_\phi) = (0, \Tilde{\theta}_\phi)$). Starting from the line with the most members as the stem, we progressively connected more lines to both sides, shifting $(\theta_r, \theta_\phi)$ of the members of the newly attached lines with suitable offsets, until all the lines are used. A more detailed explanation of the unwinding process will be presented in Panithanpaisal et al. (in prep).

Figure~\ref{fig:phases} thus allows for clear identification of the phase that each section is on, which we have also summarised in Table~\ref{tab:sections}. Although we define for both streams the section which contains the remaining progenitor (4A and 6B), we do this purely for visualisation purposes and do not make use of these sections individually. In total, discounting the progenitor-containing sections, we identify 6 apocentre sections, 7 pericentre sections and 7 intermediate sections between the two streams. The sizes of the sections vary from 30 to 318 star particles, with a total of 973 star particles within apocentre sections, 984 within pericentre sections and 1176 within intermediate sections.

\begin{figure*}
\centering
\includegraphics[width=0.99\linewidth, trim={50 0 50 30}, clip]{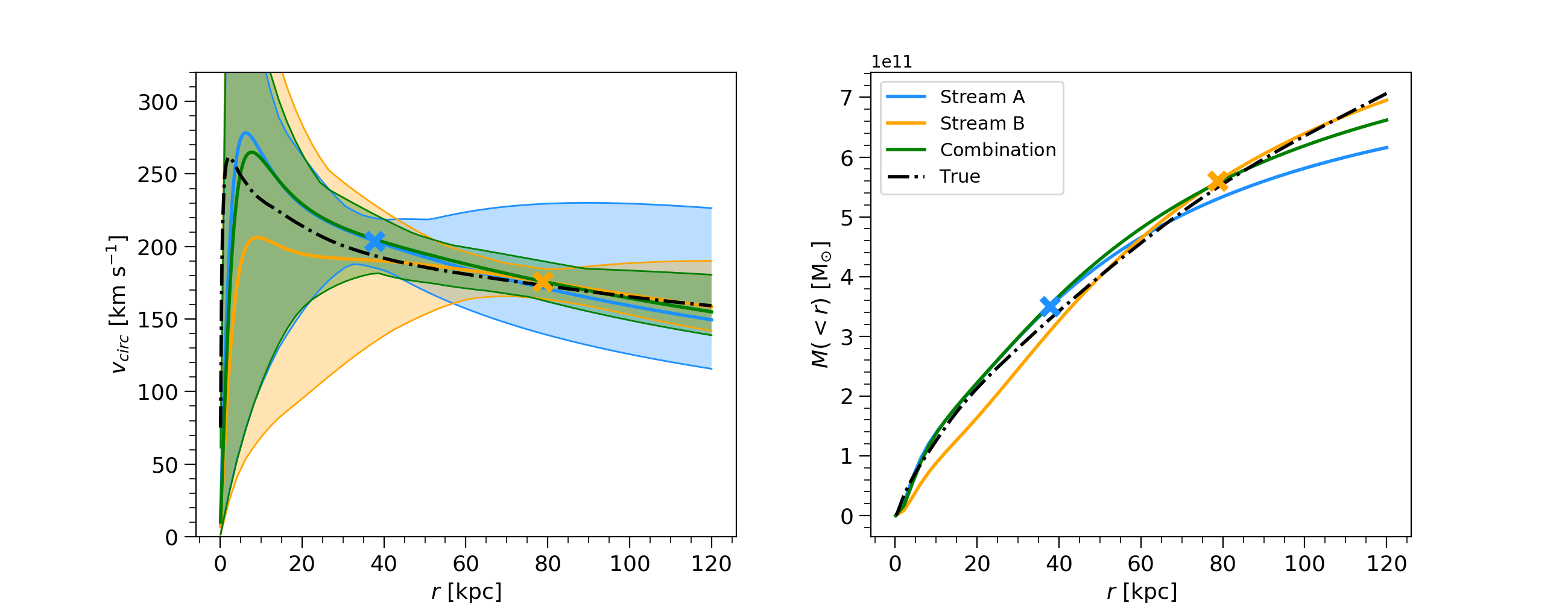}
\caption{Results for stream A, stream B and their combined data. We compare the results from these data sets to the true potential (black dash-dot line) in circular velocity and enclosed mass. The solid coloured lines show the best-fit St\"ackel potential of each data set and the shaded regions show the corresponding $1\sigma$ uncertainty regions. The median distance of the star particles in the stream A and stream B data sets are marked with a cross (see Table~\ref{tab:sections}).}
\label{fig:full}
\end{figure*}

\begin{figure*}
\centering
\includegraphics[width=0.99\linewidth, trim={0 0 0 50}, clip]{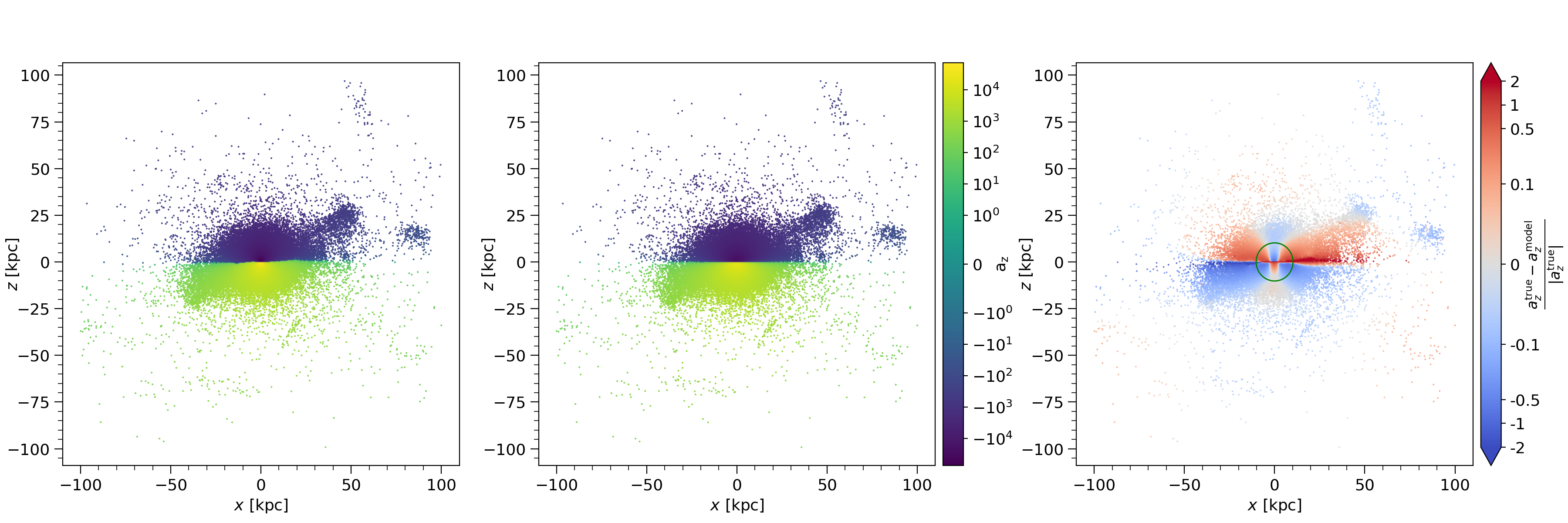}\
\caption{The binned z-accelerations of a slice of the simulated galaxy. The left panel shows the z-accelerations of the real star particles, the middle panel shows the predicted z-accelerations for the star particles based on our best-fit St\"ackel model with the combined stream data set and the right panel shows their relative residuals. The slice has a thickness of 2 kpc and is centered on galactocentric y-axis. The green circle in the right panel marks the minimum distance of the star particles in our two streams.}
\label{fig:az}
\end{figure*}

\begin{figure*}
\centering
\includegraphics[width=0.99\linewidth, trim={0 10 0 10}, clip]{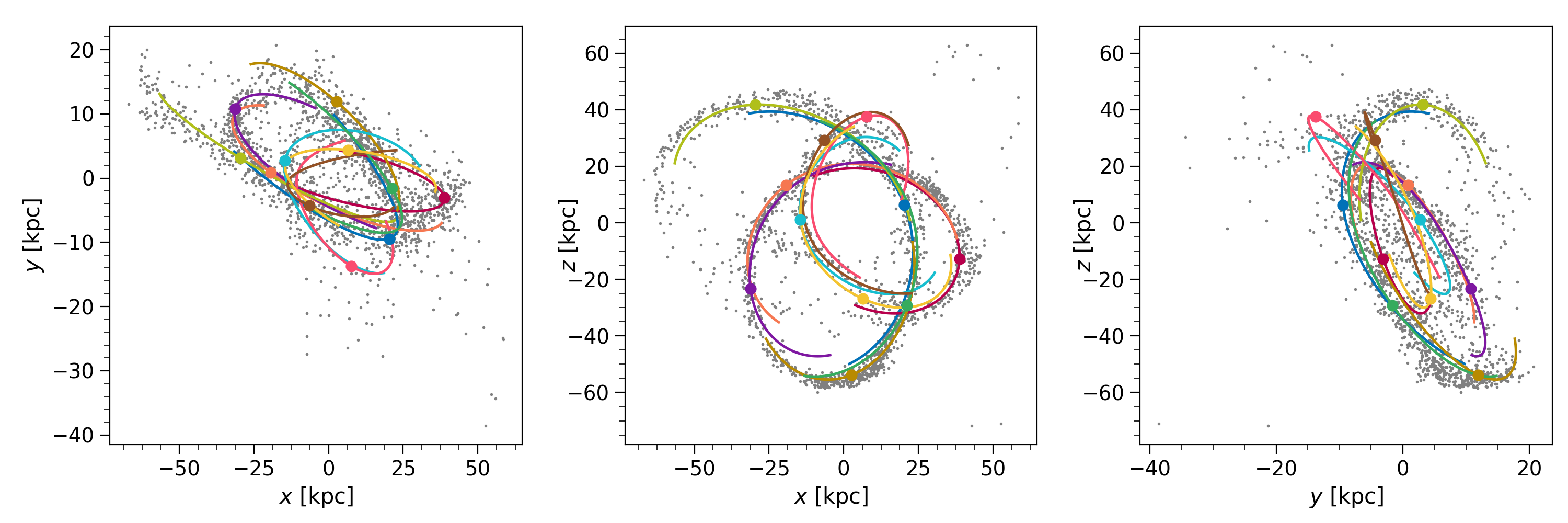}\
\includegraphics[width=0.99\linewidth, trim={0 10 0 10}, clip]{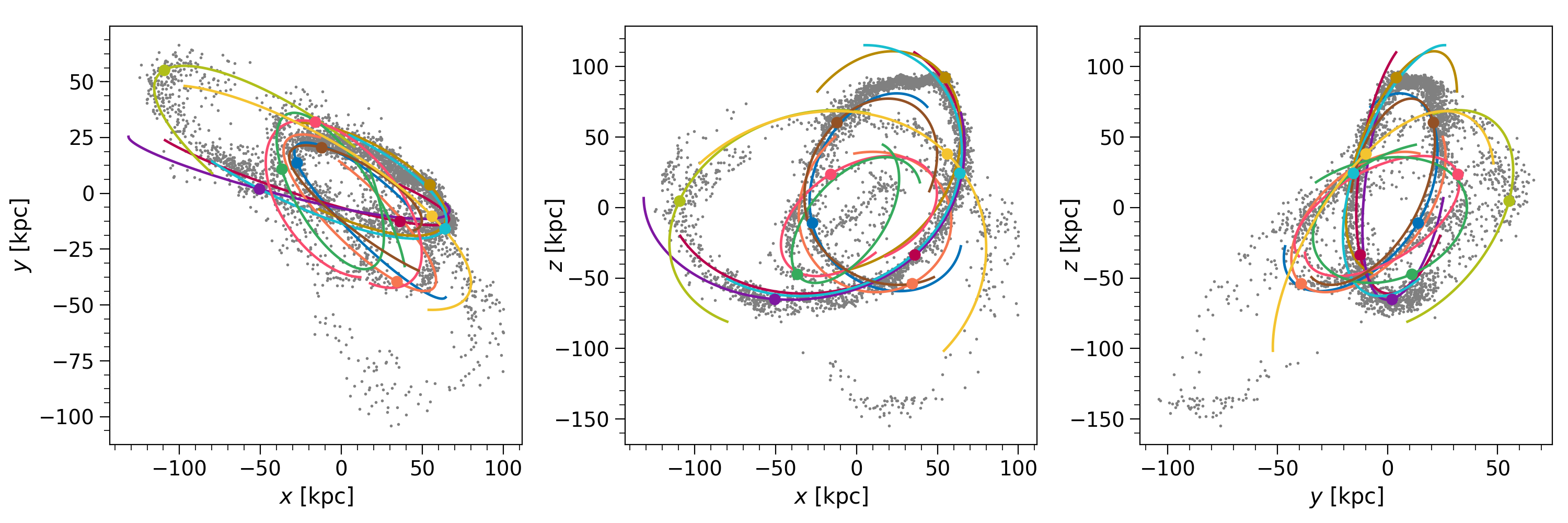}
\caption{The orbits of streams A (top) and B (bottom) in their respective best-fit potentials. The star particles whose orbits are shown (marked with a coloured dot) were picked by eye to be near the centre of their section. The orbits are coloured based on the colour scheme of the section they belong to (see Table~\ref{tab:sections}).}
\label{fig:orbits}
\end{figure*}

\begin{figure*}
\centering
\includegraphics[width=0.47\linewidth, trim={0 35 0 20}, clip]{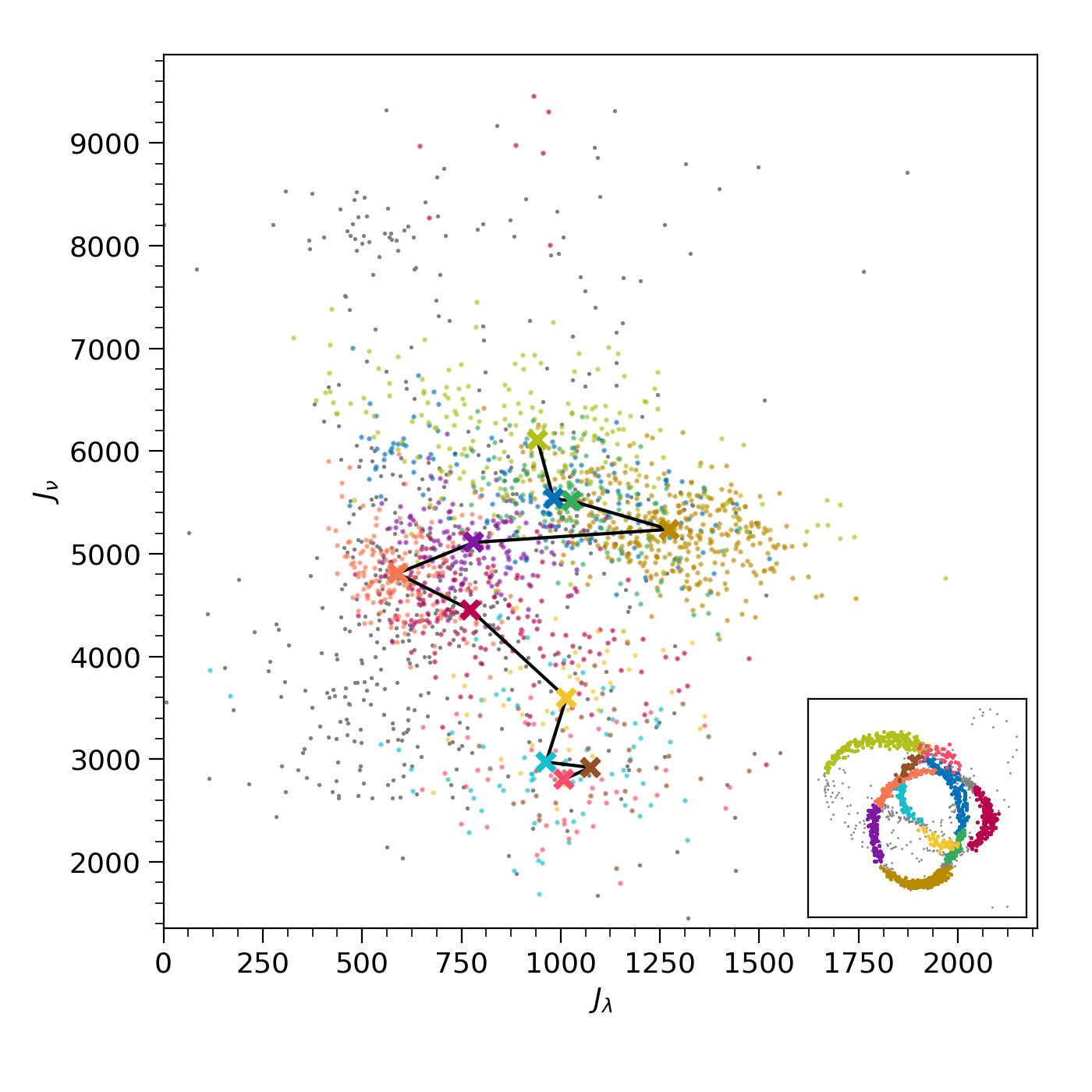}
\includegraphics[width=0.47\linewidth, trim={0 35 0 20}, clip]{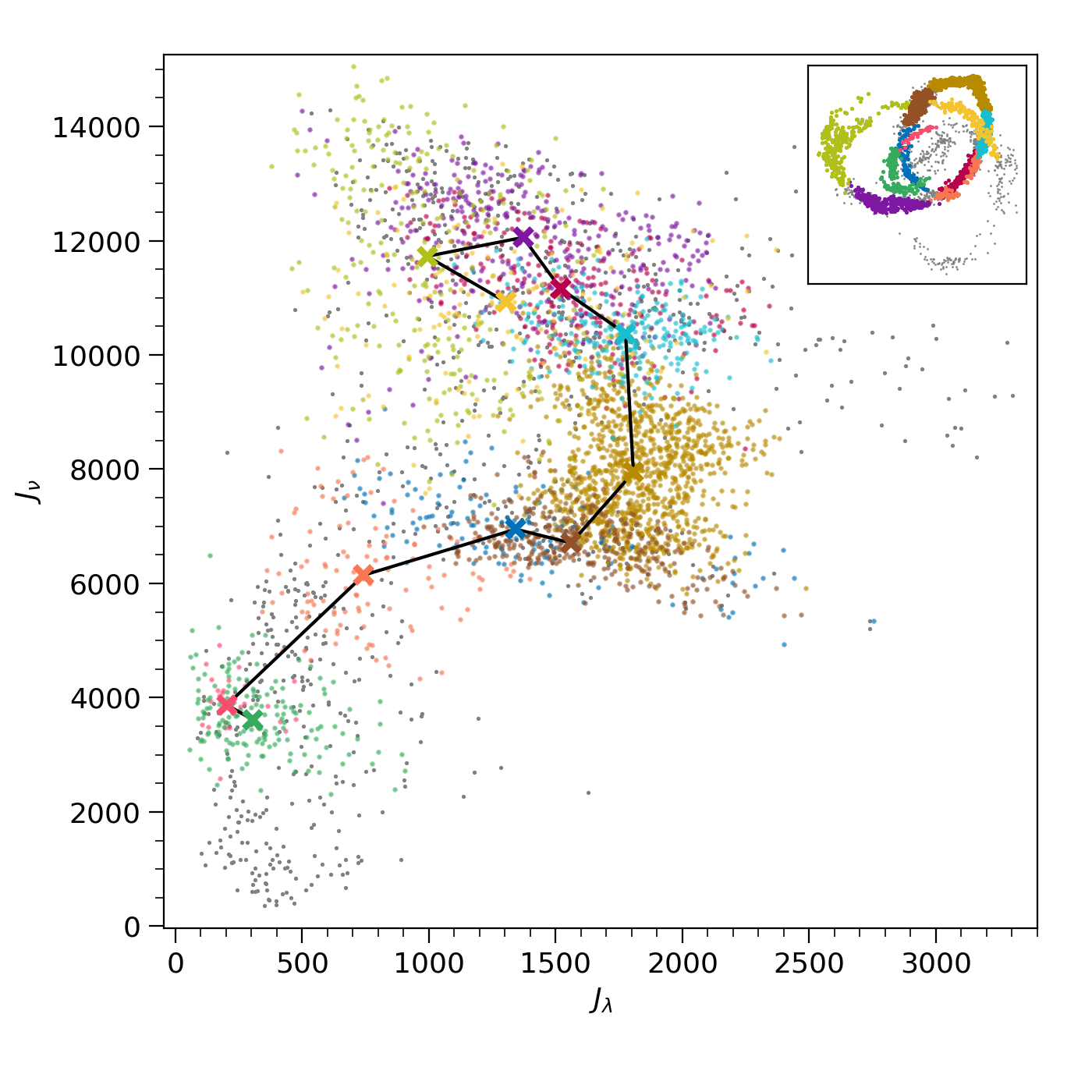}
\caption{The action-space of stream A (left) and stream B (right) in their respective best-fit potentials. The insets show the stream in x-z frame for reference. The star particles are coloured based on their sections (see Table~\ref{tab:sections}) with star particles not belonging to any section shown as grey. The centre of each section is marked with an ``X'' in its respective colour, and the black line connects them in the order that they appear in phase space.}
\label{fig:actions}
\end{figure*}

\section{Full stream results}
\label{sec:fullstreams}

In this section, we discuss the results of our action-clustering method when applied to the two streams as a whole, both independently and in combination. Figure~\ref{fig:full} compares our recovered St\"ackel potential with the true potential in two projections: the circular velocity curve (left) and enclosed mass profile (right). The true potential which is derived from the spherically binned total mass profile of the simulation snapshot is shown with the black dash-dotted line. The solid lines show the best-fit St\"ackel potential of a particular data set. Both stream A (solid green line) and stream B (solid blue line) recover the true potential relatively accurately, especially at the galactocentric distances where most of their star particles are. Stream A, whose star particles lie between $\sim 10$ and $\sim 96$ kpc, recovers the true potential more accurately at lower radii than Stream B, whose median distance is twice as large. The best-fit velocity curve lies within $32 \mathrm{\ km\ s^{-1}}$ of the true velocity curve across all the distances of its star particles, with $\sim 10\ \mathrm{km\ s^{-1}}$ at the median distance ($\sim 38$ kpc). Stream B, covering the range of distances from $\sim 20$ to $\sim 176$ kpc, on the other hand recovers the true potential better at larger radii. Its best-fit velocity curve is within $20 \mathrm{\ km \ s^{-1}}$ of the true velocity curve across its radial range, with $2.6\ \mathrm{km \ s^{-1}}$ at the median distance ($\sim 79$ kpc).
The largest differences between the predicted and true velocity curves occur at the shortest and the largest radii for both streams. The true potential at galactocentric distances less than $\sim 20$ kpc cannot be well reproduced by either stream likely due to the small number of star particles within that range (only $\sim 2\%$ of the stars in our full sample). A further discrepancy might arise also due to the lack of flexibility in the St\"ackel potential, but this possibility cannot be explored with the current data.

Unsurprisingly, the uncertainties of these best-fit measurements (shows as the shaded regions) are also the tightest where each stream contains the most data: the median distance of star particles in each stream (marked with a cross) correlates with the distance at which the uncertainty region has reduced to its minimum extent over all distances.

The results from combined data (solid green line) show improved accuracy and precision along the whole range of galactocentric distances probed by our streams. The uncertainty region no longer has a specific galactocentric distance where it is the tightest, instead we are able to recover the true velocity curve within $12\%$ over the range of radii covered by all the star particles (10 to 176 kpc). As is the case with the individual stream results, the best-fit rotation curve most deviates from the true rotation curve at very short and very large distances: between 23 to 109 kpc, the 5 to 95-percentile galactocentric distance range of the star particles, we recover the true rotation curve within $6.5\%$.

To confirm that this accuracy is typical of all streams in the simulation, we randomly selected 3 of the other 9 coherent streams present in the simulation and applied the action-clustering method to each whole stream separately. We find that all 3 streams recover the true potential very well - within 13\%, 3\% and 10\% between the 5 to 95-percentile galactocentric distance range covered by their respective star particles.

These full stream results, for which we used all the star particles of the two streams, are virtually unaffected if we remove the sections that contains the progenitors (4A and 6B). Stream B identifies the exact same potential as best-fit with or without the progenitor, while stream A finds a best-fit potential with a somewhat lower mass across all radii in the without-progenitor-section case compared to the full stream case. However, this lower mass profile is still well within the uncertainty region of both stream A and the combined streams (shown in Figure~\ref{fig:full}). The {\it extent} of the uncertainty regions between the with-progenitor and no-progenitor results only show very minor differences for both streams.

To address the question whether the difference between the best-fit St\"ackel potentials and the true potential might largely be attributed to our adoption of the St\"ackel model, we performed a least squares fit to the true velocity curve over the range of galactocentric distances where we have stellar data (10-176 kpc) using only our set of trial potentials. The potential that minimizes the least squares can be considered the best possible St\"ackel potential approximation to the true potential in terms of circular velocity. We find that this St\"ackel potential approximates the true potential extremely well: it recovers the true velocity curve within 2.8 km/s or 1.2\% over this distance range. Therefore, we expect the bias due to adopting the St\"ackel model to be negligible even for the innermost stream sections in our sample, and well within our uncertainties.

Figure~\ref{fig:az} shows the comparison of the vertical acceleration (i.e. acceleration in z-direction) field of the simulated galaxy (left panel) and that of our best-fit potential model of the combined stream data set (middle panel). The relative residuals are shown in the right panel. In general, the two acceleration fields are consistent with each other. It is not surprising that the greatest difference between the true and model z-accelerations occur near the plane of the galaxy. Part of the reason lies in the fact that the simulated galaxy is dynamic and the accelerations are not perfectly symmetric around the midplane \citep[see e.g.][]{Beane2019}. In addition to this m12i has a warp in its outer disc, that our model, or any other axisymmetric model, is not designed to reproduce. This warping is likely the cause for such pronounced differences just above the z-axis on the positive x-direction and just below the z-axis on the negative x-direction. Another reason for the mismatch lies in the fact that, like most parameterized, axisymmetric potentials, the St\"ackel model has limited flexibility when used as a global model of the potential of a realistic galaxy.
This is likely because the disc is flatter than our St\"ackel model can account for, causing it to underestimate the maximum density in the plane. In particular, as is clear from Figure~\ref{fig:full}, it does not deliver a good match to the galaxy at small radii. However, this is also the region where the uncertainties in our best-fit model are the largest, and where we have no data. The green circle on the right panel of Figure~\ref{fig:az} shows the minimum radius of the star particles in our sample.

We compare the alignment between the streams and the predicted orbits in Figure~\ref{fig:orbits}. The figure shows the orbits of a single representative star particle per each stream section, selected to lie near the centre of its respective section. These stars have then been integrated backwards and forwards in time in the best-fit potential of their respective full stream. 
We do not expect to see perfect alignment between the star particles and the integrated orbits for several reasons: there is a natural small misalignment between streams and orbits due to the range of energies in the stream stars, but more importantly in this situation, these streams have evolved is lumpy, time-evolving potentials in stark contrast to
our smooth and static St\"ackel model. The disruption of a dwarf galaxy in a realistic halo is a complex process often  taking  several  pericentre  passages,  which  each produce their own trailing and leading arms based on the evolving energy and angular momentum distribution of the stars being stripped. This naturally leads to different segments of the stream being on slightly different orbits.

Despite this caveat, the orbits of stream A (top panels) align with its star particles markedly well. The orbits of stream B (bottom panels), however, show some inconsistencies with the data. One of the main mismatches is near the location of the progenitor, which is to be expected, as leading and trailing tails by definition have a mismatch in their orbits. The second inconsistency is between the orbits of 2B and 3B. The rest of the orbits, especially those in the inner parts of the stream, align well with the data. 

The action space of each stream in their respective best-fit potentials is shown in Figure~\ref{fig:actions}. The colours of the star particles once again correspond to the sections they belong to and reveal the structure of these streams in action space. As the stars get stripped from the progenitor they settle either into leading or trailing tails, which should form two slightly separated clusters in action space. Here, we see this behaviour clearly: the section containing the progenitor (gold-coloured dots) is near the centre of the action-space and the sections comprising the trailing tail are above it, while the sections in the leading trail are below it. It is also noticeable that the sections that have evolved further from the progenitor along the tails, tend to be further from the progenitor also in action-space. This is because the least-bound stars, with the largest energy and action difference from the progenitor, escape the progenitor first.

On the whole, we conclude that the action-clustering method and the two-component St\"ackel potential can reproduce the true global potential of the simulated galaxy fairly well and without obvious biases.

\begin{figure*}
\centering
\includegraphics[width=0.95\linewidth, trim={50 0 50 33}, clip]{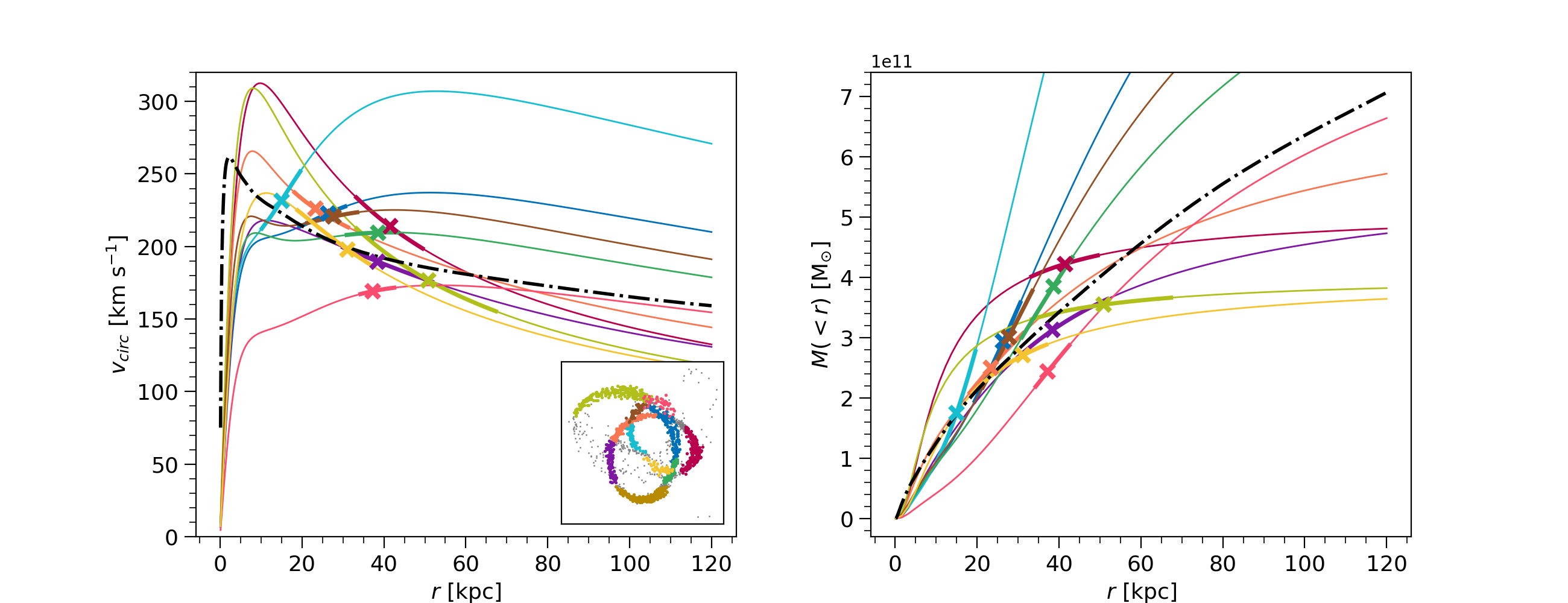}\\
\includegraphics[width=0.95\linewidth, trim={50 0 50 33}, clip]{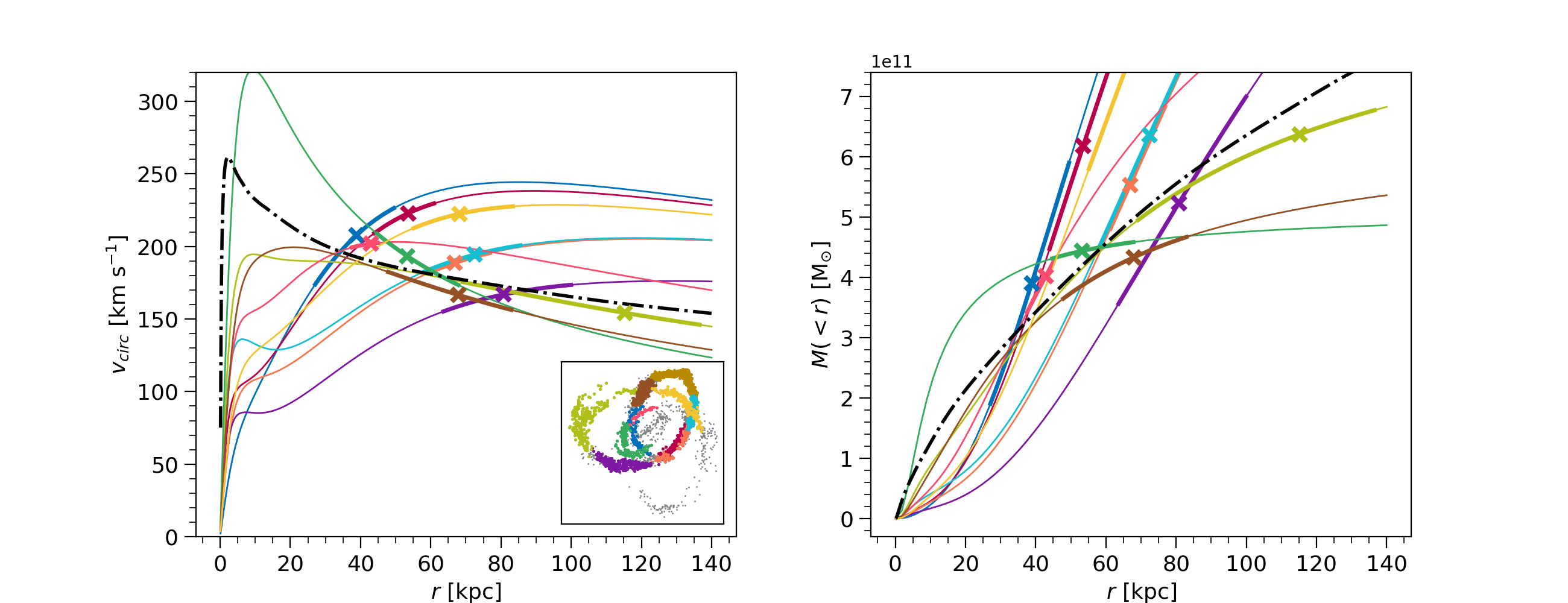}
\caption{The best-fit St\"ackel potentials for different sections of stream A (top panels) and stream B (bottom panels) shown in circular velocity and enclosed mass profiles. The crosses mark the median distance of each section, while the thicker part of the lines corresponds to the full range of distances of star particles in that section. The insets show the stream in x-z frame for reference.}
\label{fig:stream_sections}
\end{figure*}

\section{Sections of streams}
\label{sec:sections}

With simulated streams it is convenient to utilize the full stream in our analysis. However, we typically observe only the close-by segments of whole streams or, even if multiple sections are observed, we may not realise that they belong to the same larger structure \citep[e.g.][]{Bonaca2021}.

To investigate the possible consequences, we split our two streams into sections (as specified in Table~\ref{tab:sections} and Figure~\ref{fig:streams}) and use them in our analysis individually as if they were independent streams.

Figure~\ref{fig:stream_sections} show the best-fit results for all sections of stream A (top panels) and stream B (bottom panels).
There is considerable variation in the global fit of the different sections, further evidence that individual streams that span only a small region of the position-velocity space can lead to biased estimates of the global galactic potential. The majority of the sections do, however, give a good local prediction of the potential. With two exceptions, the crosses that signify the median distance of the star particles of that section, lie within $25\ \mathrm{km\ s^{-1}}$ of the true velocity curve as shown with the black dash-dotted line. 

The galactocentric radii at which each section gives the tightest uncertainties correlates linearly with their median galactocentric radius. 
The same relationship was found by \cite{Bonaca2018} who further discovered that they could tighten this correlation by adding flexibility to their potential model. In other words, the more flexible a model, the more localised the best constraints became.
This demonstrates that streams (or stream sections) do not contain information about the entire extent of their orbit, but rather are sensitive to the underlying potential at their current location. Furthermore, \cite{Penarrubia2006} showed that the past history of an evolving gravitational potential cannot be constrained using present day observables: the properties of stellar streams only reflect the present day galactic potential.

As streams recover best the current potential at their current location, the variation we see in Figure~\ref{fig:stream_sections} is unsurprising - the sections behave as if they were completely separate streams.

Finally, single streams have been shown to produce biased estimates of their host's potential \citep[see e.g.][]{Bonaca2014, Lux2013}. This serves to add even more complexity to the differences we see in Figure~\ref{fig:stream_sections}.
In the next section we will explore the underlying causes for the variations we see in both the global and local results between different streams.

\begin{figure*}
\centering
\includegraphics[width=1\linewidth, trim={40 0 40 35}, clip]{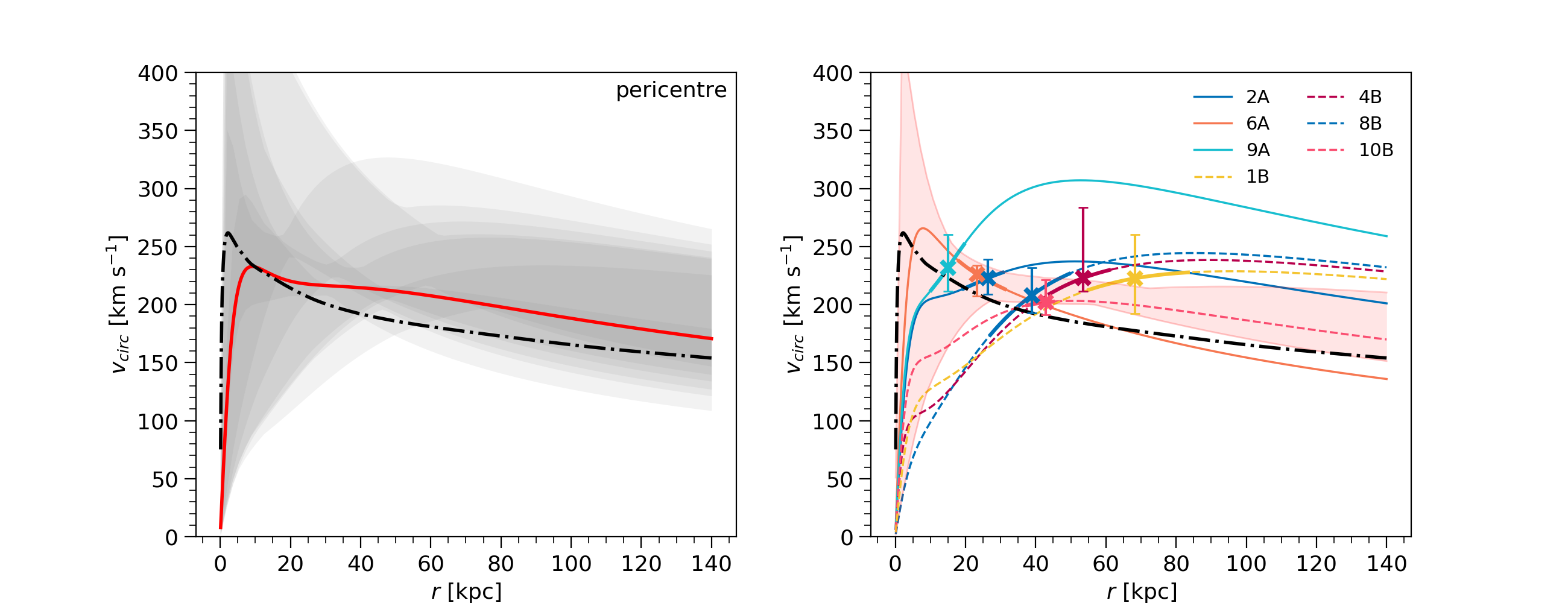}\
\includegraphics[width=1\linewidth, trim={40 0 40 30}, clip]{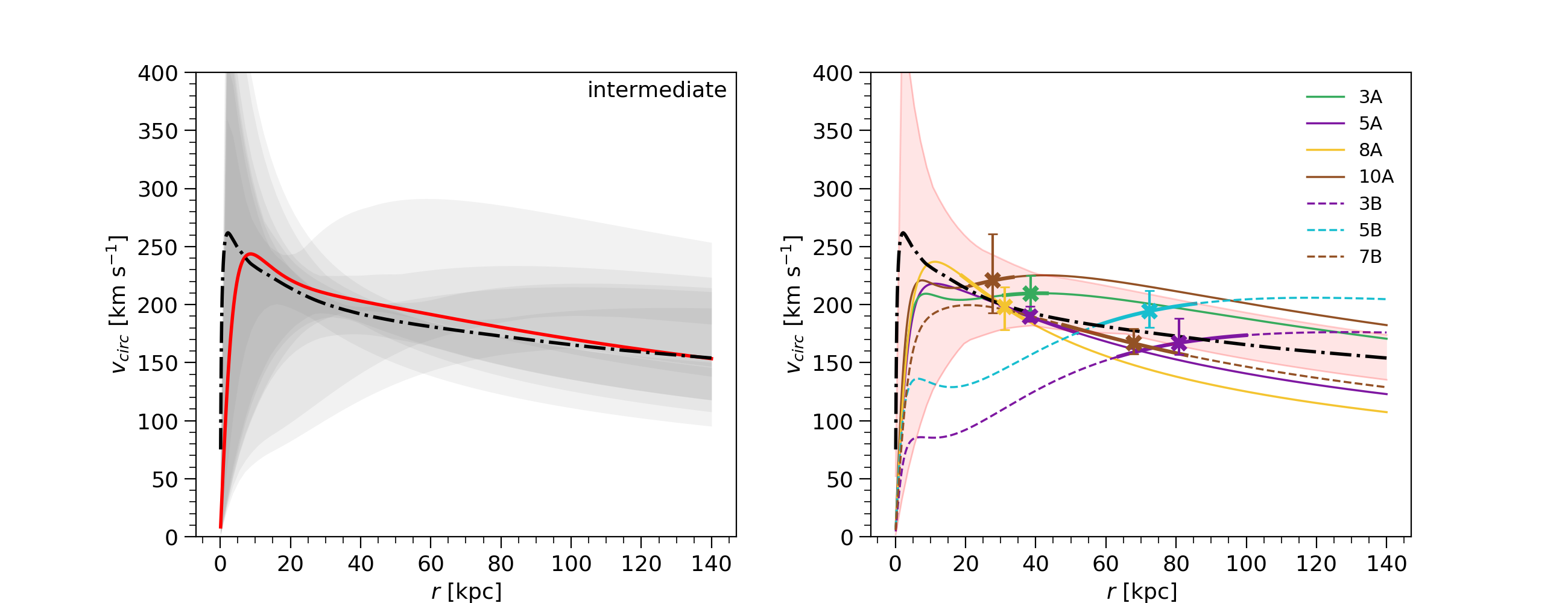}\
\includegraphics[width=1\linewidth, trim={40 0 40 30}, clip]{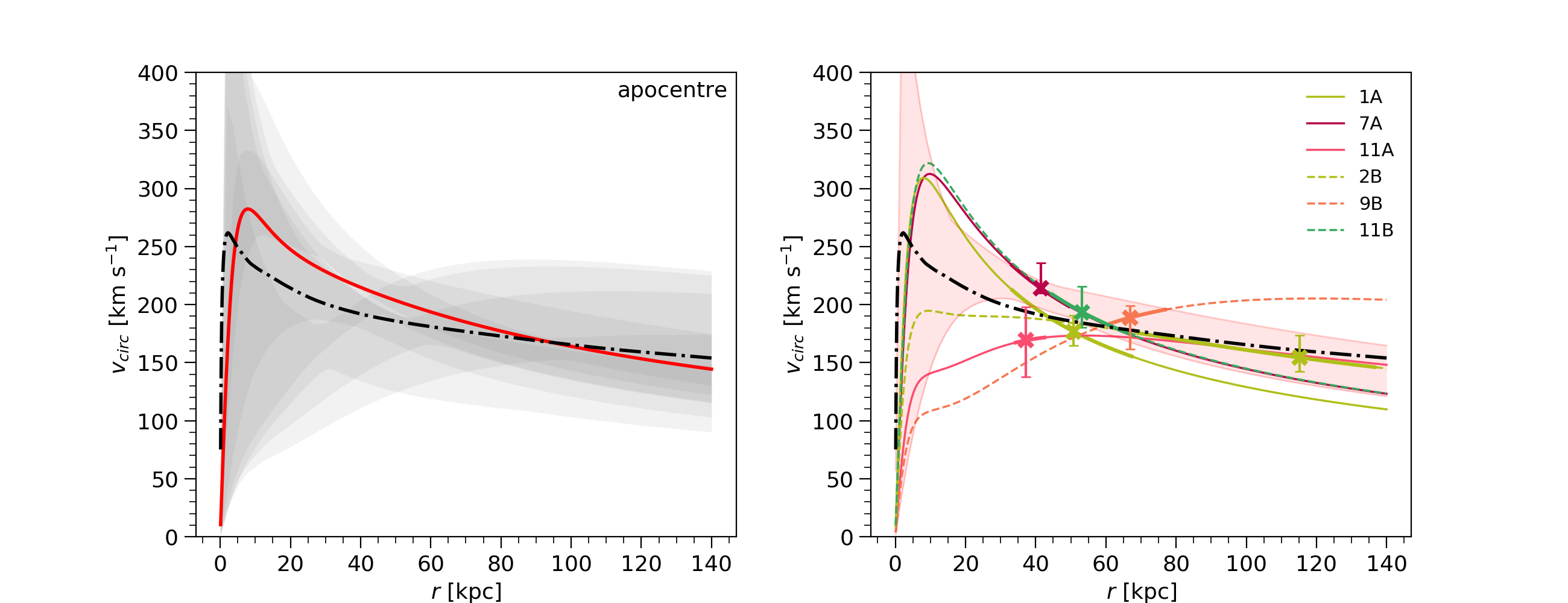}
\caption{The results of all stream A and stream B sections organised by orbital phase. The top panels show the pericentre sections, while the middle and bottom panels show the intermediate and apocentre sections, respectively. The coloured lines in the right panels show the best-fit St\"ackel potential of the individual sections, with the error bar indicating the uncertainty at their median distance. These are coloured based on Table~\ref{tab:sections}, with stream A sections shown with solid lines and stream B sections shown with dashed lines. Their full confidence regions are shown as a function of radius on the left panel as grey semi-opaque regions. The red line in the left panel shows the best-fit St\"ackel potential of the joint phase data set, and the red shaded area on the right panel is its associated uncertainty. The true potential is shown with a black dash-dotted line in every panel.}
\label{fig:phase_results}
\end{figure*}

\section{Orbital phase}
\label{sec:phase}

We now organise the sections of both streams into three groups based on their orbital phase: pericentre, intermediate and apocentre sections. This allows us to, first, explore any systematic differences in the section results based on the phase and, second, perform a joint analysis of all the sections that belong to a particular phase group using our action-clustering method. The interplay between the section results and other stream properties is explored in Section~\ref{sec:residuals}.
Our findings are summarised in Figure~\ref{fig:phase_results}, where the top, middle and bottom panels present the results for pericentre, intermediate and apocentre sections, respectively. The right panels show again the individual stream section best-fit St\"ackel potentials while the left panels show the confidence regions of these measurements overlaid in semi-transparent grey colour: the brightness of the grey colour tells us where most of the confidence regions overlap. The error bars show each section's measurement at their median distance. The left panels also contain a red line which signifies the best-fit St\"ackel potential of the joint data set of each respective phase. The confidence region for that is shown with the red shaded region in the right panel.

The pericentre sections generally overestimate the mass of their host galaxy both locally and globally, as evidenced by their best-fit potentials and the associated uncertainties. The lower edge of their confidence regions typically just about covers the true circular velocity curve at high radii, while only a few do so closer in: there is a visible gap in the coverage of the true curve between about 40 and 60 kpc. This is confirmed by the joint phase results: the best-fit potential is consistently above that of the true potential and the confidence region only barely reaches the true potential at high radii, while being somewhat above it from about 30 to 100 kpc.

The intermediate sections, on the other hand, show a much better agreement with the true velocity curve. Although there are still large variations in the individual best-fit potentials in a global sense, the local measurements are better matched. The grey shaded regions now also clearly envelope the true potential across all distance scales. The joint best-fit potential shows good agreement with the true potential nearly everywhere: we recover the true velocity curve within $6\%$ over the range of radii covered by the data (19 to 100 kpc).

Finally, the apocentre sections again generally show good agreement with the true velocity curve. The local measurements mostly agree with the true velocities, without showing a preferred bias, while the predicted velocity curves at high radii do mostly prefer lower masses when compared to the true potential. This is also clear when looking at the uncertainty regions: while at low distances the agreement between the stream sections looks fairly chaotic, at high distances, most of the confidence regions overlap slightly below the true velocity curve. The joint best-fit curve echos these individual results: at high distances we have a fairly good fit, while at low distances the difference is quite large. 

\begin{figure*}
\centering
\includegraphics[width=0.37\linewidth, trim={0 0 950 10}, clip]{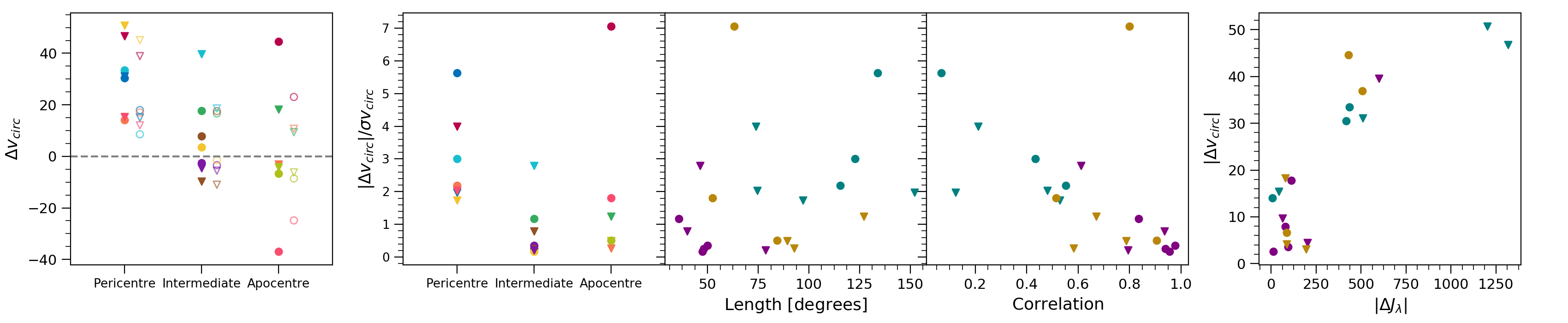}
\includegraphics[width=0.37\linewidth, trim={940 0 10 10}, clip]{Figures/residuals.png}\\
\includegraphics[width=0.9\linewidth, trim={280 0 290 10}, clip]{Figures/residuals.png}
\caption{The residuals of the circular velocity curve at the distance of the smallest $1\sigma$ uncertainty as a function of the sections' properties. On the top left panel, we also show the residuals at the sections' median distance with lighter empty markers, for comparison. On the left panels, we adopt our section-specific colour scheme, while on the rest of the panels we indicate the pericentre, intermediate and apocentre sections with teal, purple and yellow colours, respectively. A description of how the stream properties were computed is available in Appendix~\ref{sec:properties}.}
\label{fig:residuals}
\end{figure*}

\begin{figure*}
\centering
\includegraphics[width=0.9\linewidth, trim={0 70 0 70}, clip]{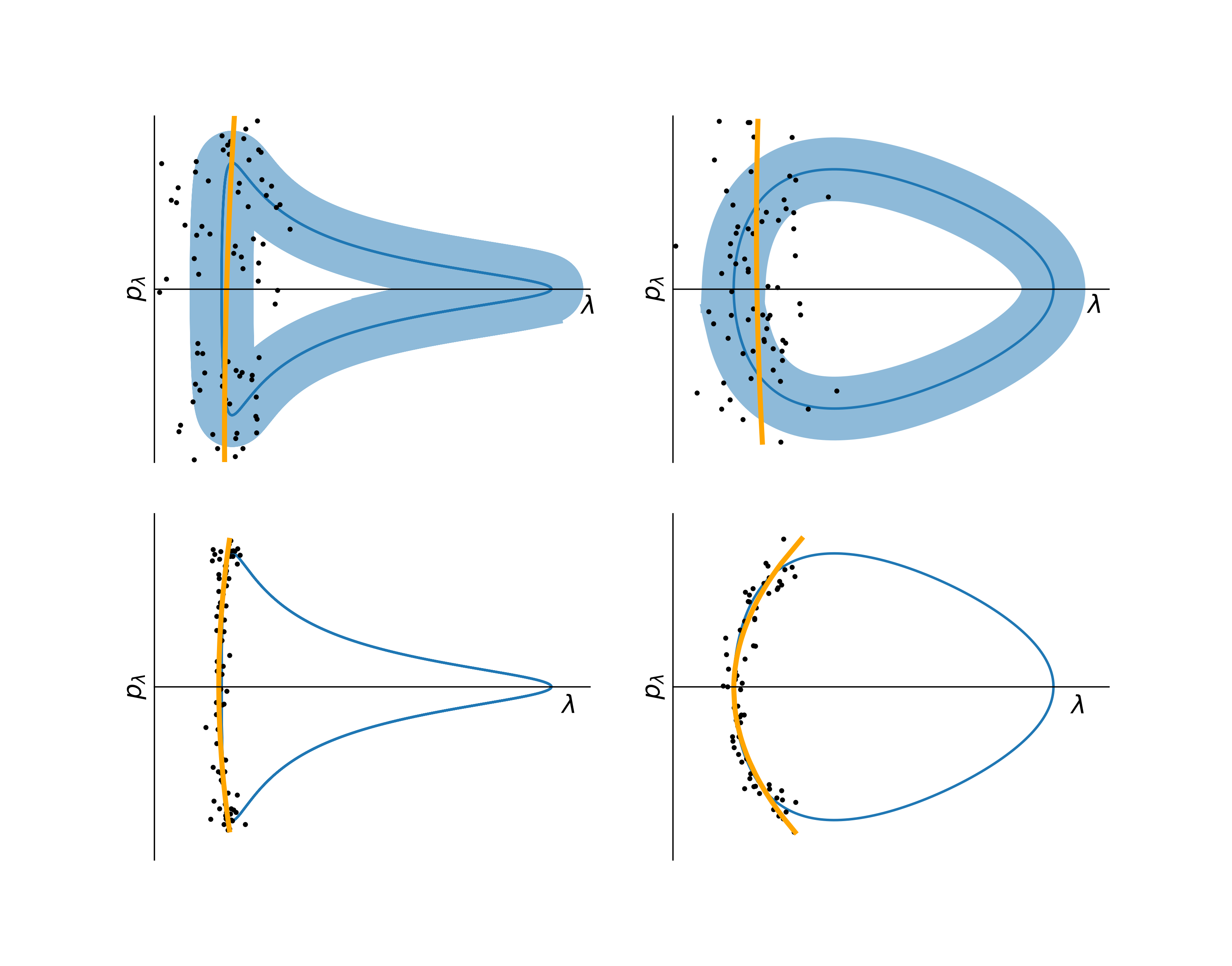}
\caption{Schematic phase diagram of dwarf galaxy streams (top panels) and cold stellar streams (bottom panels) near pericentre passage for radial orbits (left panels) and circular orbits (right panel). The black points represent stream stars created by selecting points along the pericentre of the orbit shown with the blue line and adding scatter in both $\lambda$ and $p_{\lambda}$. This scatter is 5 times larger in both positions and momenta for the dwarf streams compared to the cold stellar streams. The light blue shaded regions on top panels approximate the much larger ''scatter" in the orbits of dwarf galaxy streams. The yellow lines show a fit to the black points on each panel.}
\label{fig:cartoon}
\end{figure*}

\begin{figure}
\centering
\includegraphics[width=1\linewidth, trim={0 0 10 40}, clip]{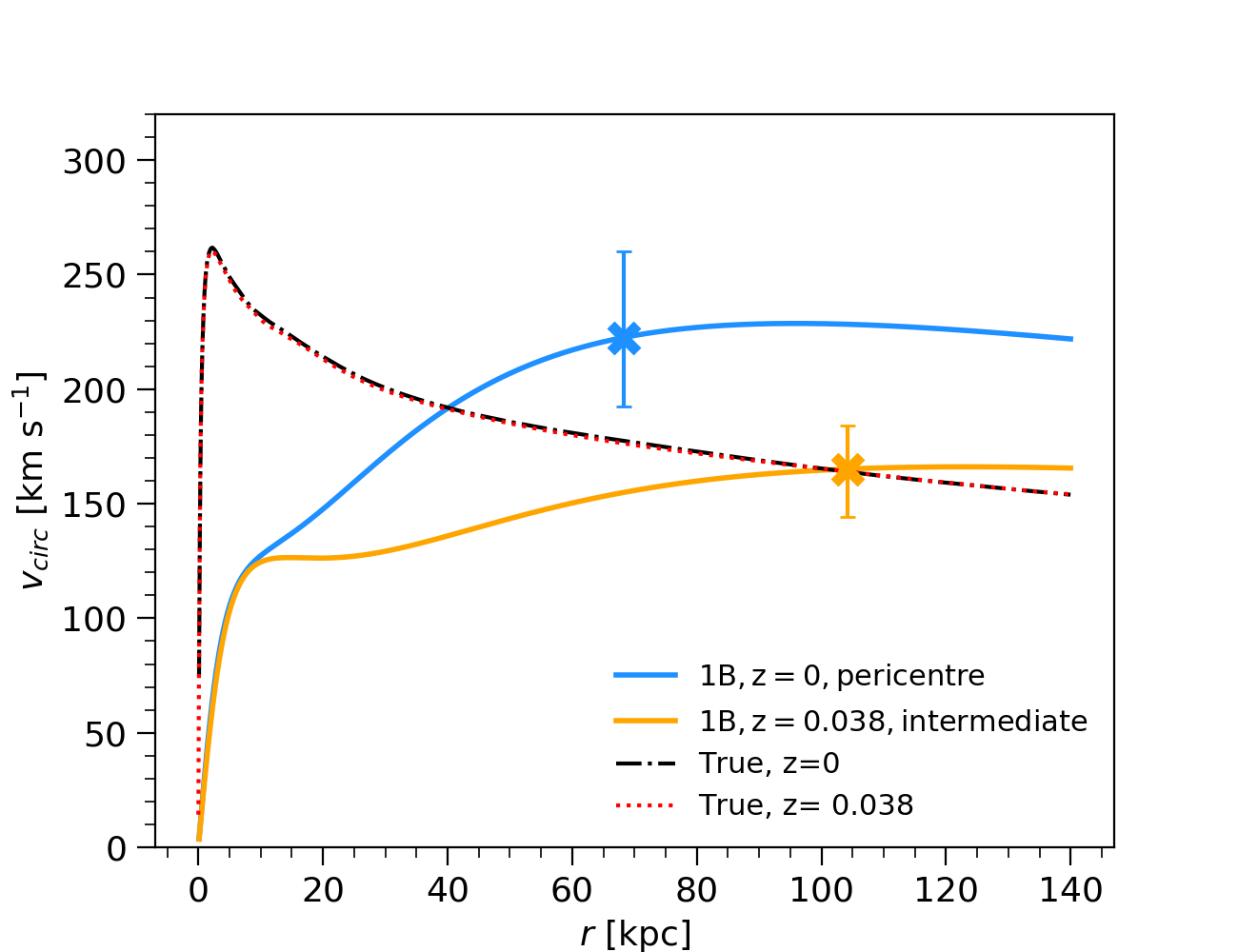}\
\includegraphics[width=1\linewidth, trim={0 0 10 40},
clip]{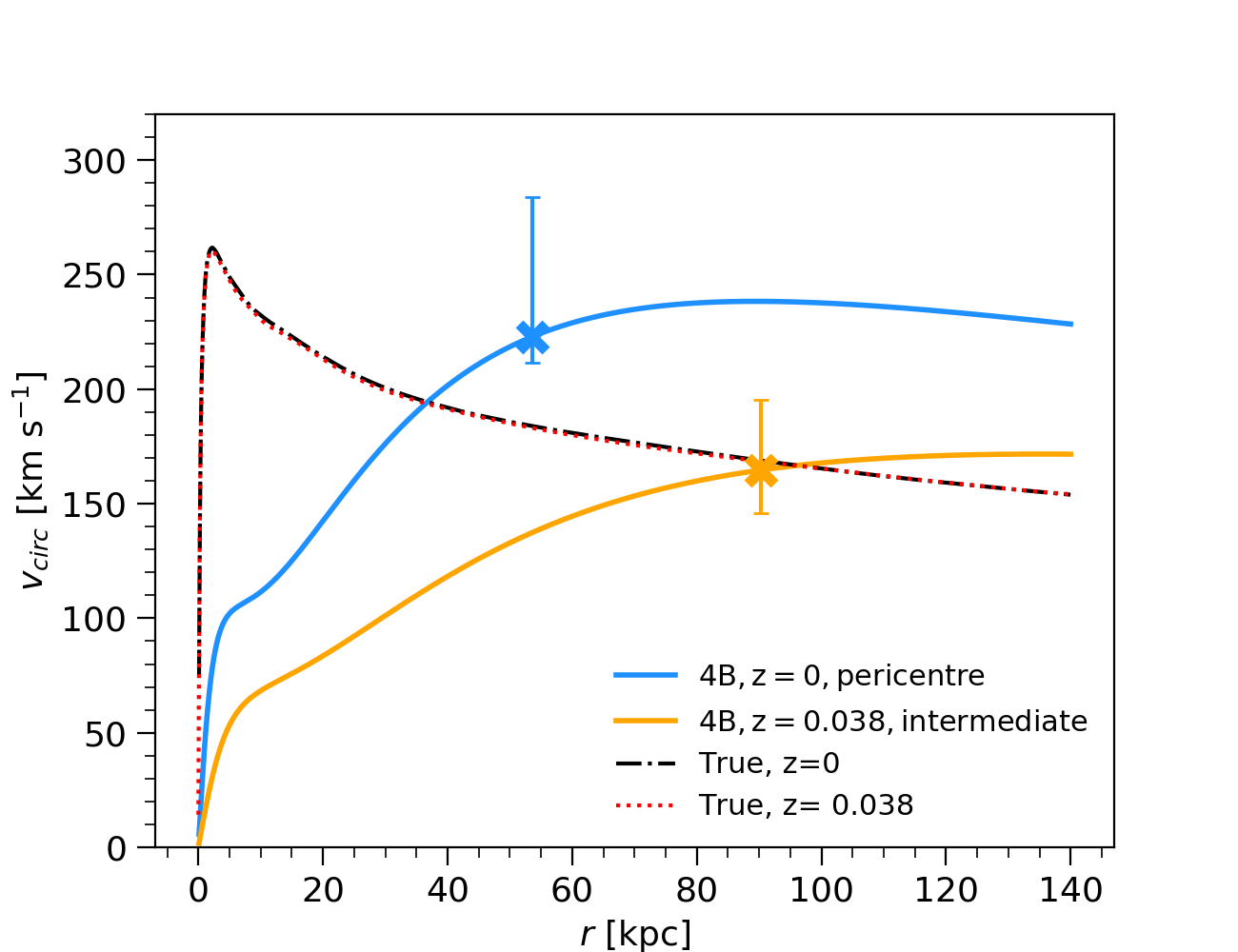}\
\caption{Comparison of the results from analysis of the present day  and past (z = 0.038) positions  of the stars in sections 1B (top panel) and 4B (bottom panel). The solid coloured lines represent the best-fit St\"ackel potential and the error bar shows the $1\sigma$ uncertainty at the median galactocentric distance for each data set. The black and red dash-dotted lines represents the true galactic potential at z = 0 and z = 0.038, respectively.}
\label{fig:past_test}
\end{figure}

\section{Bias dependence on other stream properties}
\label{sec:residuals}

In Figure~\ref{fig:residuals} we explore the dependence of the accuracy of our potential fit on several stream section properties (see also Figure~\ref{fig:residuals_appendix} where further stream properties are considered and Appendix~\ref{sec:properties} for the description of how these properties were calculated). In the top left and bottom left panels we show the points with the section-specific colours as per Table~\ref{tab:sections} while in the rest of the panels the points are coloured based on their orbital phase.
The top left panel shows for each section the residuals between the best-fit and the true velocity curve at the distance of minimum uncertainty as a function of orbital phase (solid data points). For comparison we also show, with empty markers and lighter colours, the residuals at the median distance of each section (this corresponds to the vertical difference between each of the crosses and the dash-dotted line in Figure~\ref{fig:stream_sections}).
In the bottom left panel, we show the residuals relative to their $1\sigma$ uncertainty. 

We find that there is a clear correlation between the accuracy of the fit and the orbital phase of the stream. The best-fit results of the intermediate sections show the least amount of scatter around the true velocity curve, and do not appear to have a preferred bias. The apocentre sections, although not exhibiting an obvious bias either, have a lot more scatter around the true potential. The pericentre sections, in contrast, consistently overestimate the mass. 
Moreover, their uncertainties are small compared to the residuals, indicating a systematic bias.

We do not see any clear trends with either the angular length (bottom middle panel) or physical length (shown in Figure~\ref{fig:residuals_appendix}) of the stream sections. Indeed, the intermediate sections in our sample tend to be quite short and nevertheless give better estimates than the often longer pericentre sections. 
We also see no correlations with the number of stars in each stream section, the median galactocentric distance, the median distance from the galactic plane, width or velocity dispersion in each section (all shown in Figure~\ref{fig:residuals_appendix}) nor with the angular momentum vector or the galactocentric distance range covered by a certain stream section (not shown). All of this gives us confidence that it is truly the effect of the orbital phase that causes the variation in the quality of our constraints.

To explain this effect, we investigated the correlation between $\lambda$ and $p_{\lambda}$ in each stream section. The bottom right panel in Figure~\ref{fig:residuals} shows that stream stars exhibit strong correlations between their motions and positions during the intermediate phase. The pericentre streams meanwhile have the weakest correlations. In general, the stronger the correlations between motions and positions of stream stars the better the constraints from that stream tend to be. Figure~\ref{fig:cartoon} illustrates this concept. On the upper panel, we show a cartoon of a dwarf galaxy stream near pericentre passage for a more radial orbit (top left panel) and a more circular orbit (top right panel). The black points represent stream stars, which have been created by selecting points along the pericentre of a single orbit (the blue line) and adding scatter in both $\lambda$ and $p_{\lambda}$. The light blue shaded regions approximate the ``scatter'' in these orbits. The yellow line shows a fit to the black points and is in both cases almost perfectly vertical, signifying no correlation between $\lambda$ and $p_{\lambda}$. This in turn indicates lack of statistical power for significant differentiation between different potential models.

The origin of the systematic error that we see arising with pericentre streams therefore lies in this lack of correlation between the positions and momenta of stream stars during pericentre passage. As a consequence this data cannot differentiate between potentials as successfully as intermediate and apocentre streams. In the case of the action-clustering method this property of the pericentric motion manifests as overestimation of mass. Due to the lack of correlations between motions and positions, it is possible to find a high mass potential that confines all the stream stars onto orbits with very little radial motion. This potential is then selected as the best-fit because it forms a dense cluster near $J_{\lambda} \sim 0$. However, to accommodate this configuration the stars have to be placed on a variety of different phases on their respective orbits. This means that the ordering of the stars along the stream, and in fact the spatial coherence of the stream itself, breaks down in this potential \citep[see also][]{BuistHelmi2015}. We call this spurious reordering of the stream stars in the incorrect potential ``phase scrambling''.

Although the pericentre streams are likely to yield potentials that are highly biased, the confidence regions can nevertheless be small. This is due to the fact that to calculate the uncertainty, we compare the action-space distribution of the best-fit potential to that of all other trial potentials, and draw the error contours so that they mark the boundary where the action-distributions begin to significantly differ. The weakness of this technique is that even if there were other potentials that produced a highly clustered action-space (i.e. had a high KLD1) there is no guarantee that they would be included in the uncertainty region if their action-space looks significantly different from that of the best-fit, e.g. when clusters simply form at a sufficiently different locations in action-space. So our set-up is reliant on having managed to determine the ``correct'' potential as the best-fit, while the uncertainty just measures the variation around it.

Finally, on the top right panel, we show the goodness of fit as a function of the difference in median $J_{\lambda}$ between the best-fit potentials of individual sections and that of the combined full streams (green line in Figure~\ref{fig:full}). The further the individual results are from the true potential, the bigger this change in $J_{\lambda}$. Although it is expected that a greater difference between two potentials results in a greater change in the action-space, we do not see such a trend with $J_{\nu}$.

To confirm that the quality of our results is indeed determined by the phase of the stream, we looked at the two stream sections with the highest difference between the true and estimated velocity curves -- pericentre sections 1B and 4B -- in a simulation snapshot corresponding to redshift z = 0.038, when both of these stream sections occupied the intermediate phase. We reapplied the action-clustering algorithm for the stars in each section but now using their past positions and velocities from snapshot z = 0.038. The results of this test are presented in Figure~\ref{fig:past_test}, where with the blue line we show the result of the analysis of the current day (pericentre phase) positions of the stream stars, and with the yellow line the past (intermediate phase) positions of the same stars. In both cases, the results originating from the past positions, when the stars were at intermediate phase, perform markedly better.

\section{Conclusions and Discussion}
\label{sec:discussion}

In this work, we have shown that with two {\it whole} dwarf galaxy streams we recovered the true rotation curve of the simulated galaxy within $12\%$ over the entire range of radii covered by our set of star particles (10 - 176 kpc) when adopting the two-component St\"ackel model. However, this accuracy is much improved over the distances where we have the most data, recovering the true rotation curve within 6.5\% between the 5 to 95-percentile distance range (23 - 109 kpc). This leads us to conclude that using the St\"ackel potential does not introduce a significant bias into our results, at least not more so than any other axisymmetric, parameterized potential model would.

\cite{Bonaca2014} explored the accuracy of smooth analytic potentials in representing realistic - lumpy and time-evolving - dark matter halos. Using a collection of streams evolved (using the streakline method) in the Via Lactea II simulation, they estimated the Galactic mass by comparing these “observed” streams to models generated in trial analytic potentials and showed that just assuming an analytic potential limits the measurement accuracy to $5-20\%$. This limit was reached only with the full collection of 256 streams in their sample, while individually the streams were much less accurate (only $40-60\%$ of the individual streams could recover the true parameters within $10\%$). 
A similar exploration was performed by \cite{Sanderson2017} who fitted analytic potentials to the streams occurring natively in the Aquarius A simulation using the action-clustering method.\footnote{Although mostly similar to the action-clustering method described here \cite{Sanderson2017} used the product of the marginal distributions of $p$ instead of the uniform distribution as the comparison distribution $q$, gave equal weight to all stars and did not separate different streams during the process of density estimation.} They found that with simultaneous fitting of 15 streams they could recover $M_{200}$ within $10\%$.

The tight constraints we achieve here with just two streams are remarkable considering that, in contrast to Via Lactea II and Aquarius simulations, both of which are dark matter-only simulations, we model a galaxy from a fully cosmological-baryonic simulation which contains a stellar and gas disc shaped by star formation in addition to a time-evolving dark matter halo. Yet we obtain this precision using only a global, two-component St\"ackel potential to represent the entire complexity of this galaxy. This is likely due to the good orbital phase coverage of these two streams: both streams have several wraps around the host galaxy and as such cover each orbital phase multiple times.

Both \cite{Bonaca2014} (using streakline) and \cite{Sanderson2017} (using action clustering) demonstrated that an oversimplification of the potential model does not intrinsically produce a biased mass profile when fitting a collection of streams. Our findings agree with this: our results with the full streams show no presence of systematic bias (Section~\ref{sec:fullstreams}). 

We next split each of the two streams into 11 smaller sections based on their orbital phase (Section \ref{sec:sections}) and analysed them independently. We find that the quality of the constraints on the mass profile depends on the orbital phase of the stream (Section \ref{sec:residuals}).
There is a clear systematic bias when using only the pericentre streams in our analysis: this data overestimates the mass of the host galaxy at all galactocentric radii (see Figure~\ref{fig:phase_results} top panel and Figure~\ref{fig:residuals} lower left panel). This systematic error stems from the fact that during pericentre passage the positions and momenta of stream stars are not correlated (see Figure~\ref{fig:residuals} bottom right panel). Although a joint fit of multiple streams is usually recommended to get a better fix on the potential, this bias remains even when all pericentre sections are analysed jointly. We find that streams on the intermediate phase are the most likely to give bias-free local mass estimates individually, and a bias-free and accurate global mass profile in combination (see Figure~\ref{fig:phase_results} middle panel and Figure~\ref{fig:residuals} left panels). 

In \cite{Reino2021} we showed that when analysed with the action-clustering method GD-1 data produced a mass estimate that was considerably larger than those from Pal 5, Orphan and the combination of all three streams. We explored the range of orbital phases the GD-1 stars were on with the best-fit GD-1 potential and found that the stars were all placed on very different orbital phases on their respective orbits. We briefly discussed that the cause for this, and therefore the high mass that GD-1 recovers, is likely due to the natural energy gradient along the stream not being reproduced. This is another symptom of the phase scrambling we discussed above and, since GD-1 is believed to be a pericentre stream, it aligns with our results here.

This inability of pericentre streams to distinguish robustly between potential models can manifest in other ways for different methods. Previously, \cite{SandersBinney2013} remarked on having more difficulty constraining the potential with streams observed at their pericentre.
They analysed a mock stream with their angle-frequency slope method both during its apocentre and the subsequent pericentric passage and found that they could not recover the true potential parameters as successfully in the pericentric case. However, despite several local minima in their likelihood surface, they do not calculate a large systematic bias. The cause for such a behaviour could be that streams cover a smaller angle space near their pericentres making the determination of the slope more prone to errors.

Similarly, \cite{Koposov2010} found that they could not constrain all the parameters in their 3-component potential model with 6-dimensional GD-1 data using the orbit-fitting method and noted that, due to being near its pericentre, GD-1 might not have sufficient phase coverage to differentiate between orbits produced in different potentials, resulting in a poor fit.

Pericentre streams therefore lead to either biased results or weak constraining power irrespective of the applied method. Although our position in the Galaxy and the increased stellar density of streams near pericentre makes detecting streams near their pericentres the easiest, their relative accessibility does not lead to an appreciable improvement in our understanding of the Galactic potential or the field of near-field cosmology.

Although in this work we have made use of dwarf galaxy type streams, the conclusions drawn here are also applicable to most cold stellar streams of globular cluster origin. This is because the cause for the weak constraining power in pericentre streams is not unique to dwarf type streams. In the bottom panels of Figure~\ref{fig:cartoon} we show the cartoon versions of two cold stellar streams to provide comparison with the dwarf galaxy streams in the top panels. Both cartoon streams in the left panels were created from the same original orbit (the blue line), however, the cold stellar stream stars are 5 times less scattered both in $\lambda$ and in $p_{\lambda}$. The same holds true for the streams in the right panels: the stars were created from the same original orbit, but with 5 times more scatter added to the dwarf galaxy stream stars. Although with considerably less scatter, the positions and momenta of stars in cold stellar streams on more radial orbits (bottom left panel) would still be uncorrelated during pericentre passage and therefore result in poor constraining power. However, in contrast with dwarf galaxy streams, a cold stellar stream on a more circular rather than radial orbit (bottom right panel) can potentially have sufficient curvature in the pericentre part of the phase diagram to constrain a model potential.

Many studies have shown that in general longer streams have more constraining power. When investigating the information content in the tracks of stellar streams \cite{Bonaca2018} found that longer streams (in degrees) achieve the highest precision in recovering the potential parameters. We do not find any correlation between either angular or physical distance with the accuracy of the constraints or the precision of our confidence regions. In fact, our intermediate phase streams often tend to be the shortest and the pericentre streams the longest. However, our shortest streams are around $\sim 50$ degrees while only one of the 11 streams \cite{Bonaca2018} studied reaches this length, the rest being considerably shorter. It could be that the trend with length is no longer as relevant as other factors when it comes to longer streams. Alternatively, the effect could be related to the difference in our methods. \cite{Bonaca2018} made comparisons between the tracks of stream data and models in position and velocity space, so longer streams will allow the comparisons to be made over a larger extent and thus enhance the results. Conversely, the length of the stream has no direct impact on the constraints derived with the action-clustering method as we are only measuring the density of the stars in action space. 

We summarize our findings as follows:
\begin{itemize}
    \item Although individual streams are likely to deliver accurate estimations of the local galactic profile, they should not be relied on for yielding good global fits.
    \item We have shown that the pericentre streams can lead to significant systematic errors when used to constrain the potential of their host galaxy. 
    \item Meanwhile apocentre and, especially, intermediate phase streams lead to accurate inference. 
\end{itemize}

For accurate high confidence constraints on the Galactic potential we therefore advocate targeting streams that are likely at intermediate or apocentre phases. 

\section*{Acknowledgements}

SR would like to thank the Faculty of Science of Leiden University for their support through the Leiden/Huygens fellowship.
RES and NP gratefully acknowledge support from NASA grant 19-ATP19-0068. RES is additionally supported by HST grant AR-15809 from STScI, NSF grant AST-2007232, and a Scialog-Time Domain Astronomy Award from the Research Corporation for Science Advancement. This work was performed in part at Aspen Center for Physics, which is supported by National Science Foundation grant PHY-1607611.

Simulations used in this work were run using XSEDE supported by NSF grant ACI-1548562, Blue Waters via allocation PRAC NSF.1713353 supported by the NSF, and NASA HEC Program through the NAS Division at Ames Research Center. The authors also would like to thank the Flatiron Institute Scientific Computing Core for providing computing resources that made this research possible, and especially for their hard work facilitating remote access during the pandemic. Analysis for this paper was carried out on the Flatiron Institute's ``Iron'' computing cluster, which is supported by the Simons Foundation.

\section*{Data availability}
The data of the two streams used in this manuscript are available as part of the SAPFIRE: Streams on FIRE catalogue at \href{https://flathub.flatironinstitute.org/sapfire}{https://flathub.flatironinstitute.org/sapfire}.

\appendix
\section{Stream velocities}
\label{sec:stream_velocities}

\begin{figure*}
\centering
\includegraphics[width=0.97\linewidth, trim={0 5 0 10}, clip]{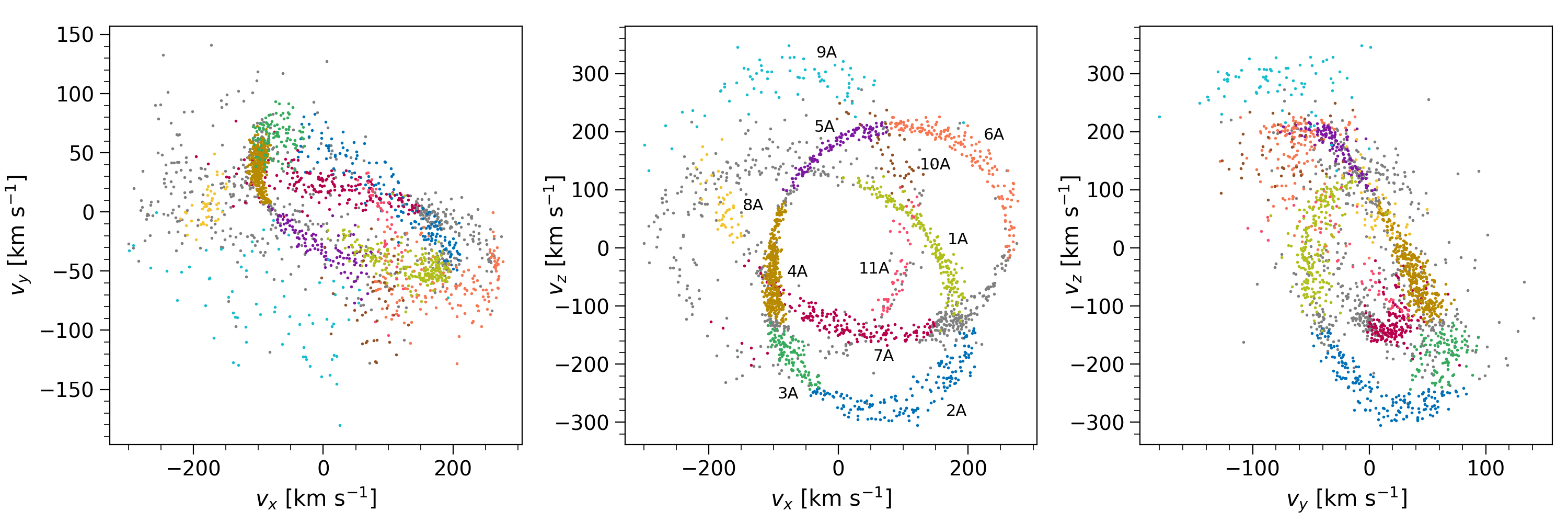}\
\includegraphics[width=0.97\linewidth, trim={0 5 0 10}, clip]{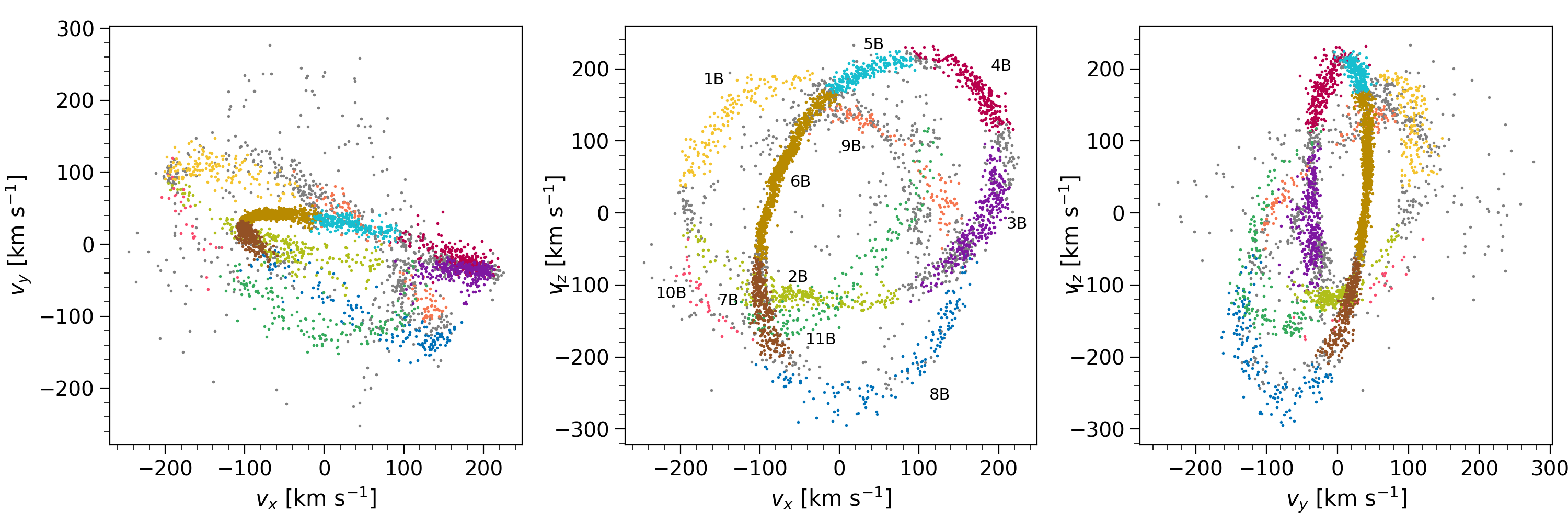}
\caption{Galactocentric velocities of streams A (top) and B (bottom). Each stream section is identified by the colour given in Table~\ref{tab:sections} and a label. Star particles not belonging to any section are shown as grey.}
\label{fig:streams_v}
\end{figure*}

In Figure~\ref{fig:streams_v} we show the galactocentric velocities for our two streams with each section highlighted in the colour as specified by Table~\ref{tab:sections}.

\section{Bias as a function of further stream properties}
\label{sec:properties}

\begin{figure*}
\centering
\includegraphics[width=0.99\linewidth, trim={50 60 20 363}, clip]{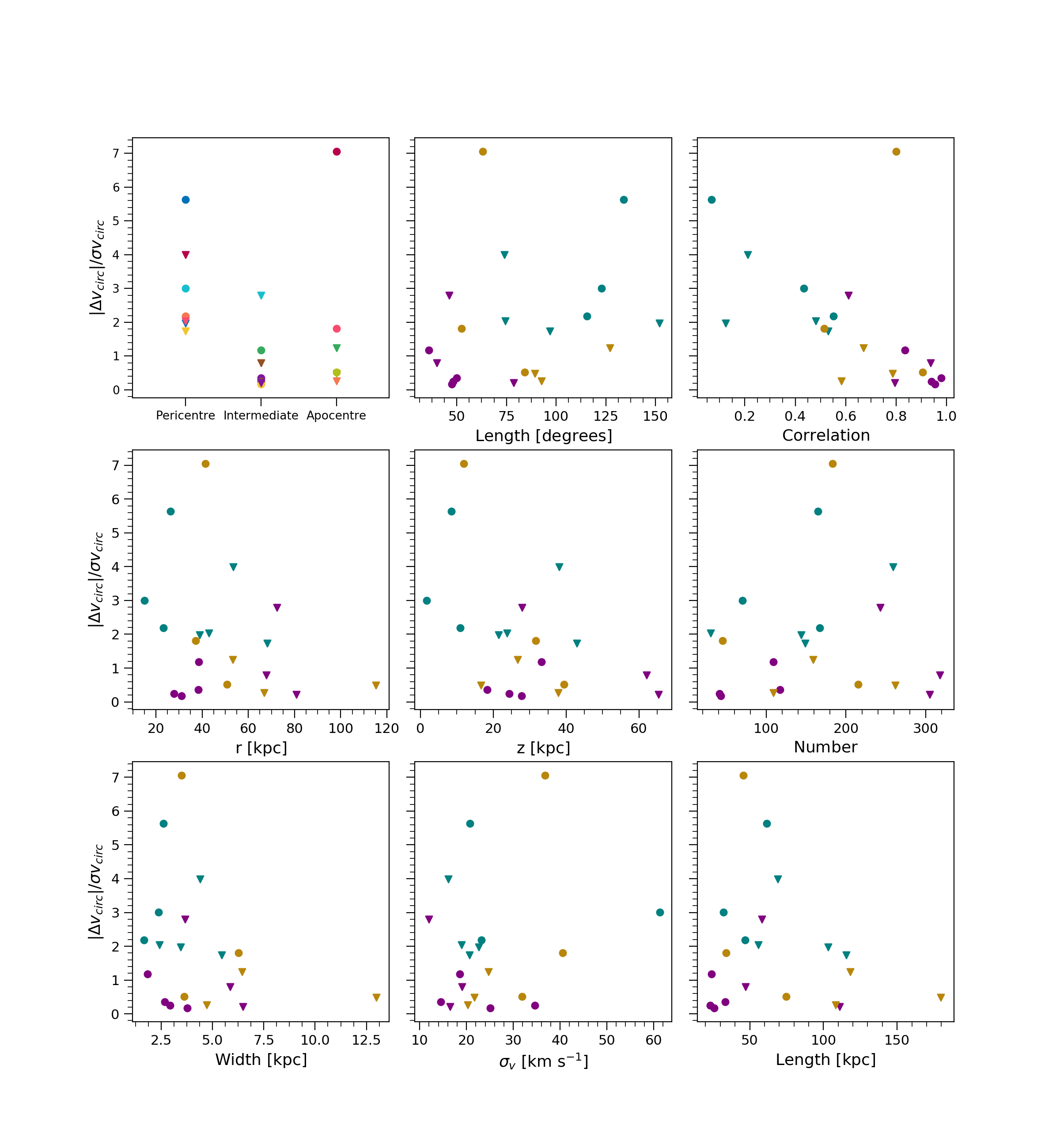}
\caption{The residuals of the circular velocity curve at the distance of the smallest $1\sigma$ uncertainty as a function of the sections' properties. We indicate the pericentre, intermediate and apocentre sections with teal, purple and yellow colours, respectively.}
\label{fig:residuals_appendix}
\end{figure*}

In this section, we show in Figure~\ref{fig:residuals_appendix} the accuracy of our potential fit as a function of some further stream section properties. As already remarked in Section~\ref{sec:residuals}, none of these stream section characteristics are sufficient to explain the variation that we see in the quality of our potential constraints.

To measure the stream length, width and velocity dispersion we first convert to stream-aligned coordinates. This is a spherical coordinate system $(r, \xi,\eta)$ where the equator, $\eta = 0 $, is defined by a great circle best-fitting the stream data centered on the the galactic centre and the radius, $r$ is the median galactocentric radius of the stream stars.
The angular length of the stream is then defined as $\Delta \xi$. The physical length is found by computing the arc length of the circle subtended by the stream, i.e. $r \times \Delta \xi$. To estimate the width of the stream, we fit the galactocentric x, y, and z coordinates of stream stars as a function of the angle along the stream, $\xi$, with a quadratic polynomial. The distance of each star from this stream "axis" can then calculated at their respective $\xi$, i.e. $d_i = \sqrt{(x_i - x(\xi_i)^2 + (y_i - y(\xi_i)^2 + (z_i - z(\xi_i)^2 } $. We then define the width as the root-mean-square of these distances. An analogous technique is employed to calculate the velocity dispersion, $\sigma_v$, except now the polynomial is fitted to the galactocentric $v_x, v_y, v_z$ as a function of $\xi$.
Correlation between $\lambda$ and $p_{\lambda}$ for each stream is defined as the absolute value of the Pearson correlation coefficient between these coordinates.

\bibliographystyle{mnras}
\bibliography{fire_streams} 

\bsp	
\label{lastpage}
\end{document}